\documentclass[]{aa}  

\usepackage{graphicx}
\usepackage[colorlinks=true, linkcolor=blue, citecolor=blue, filecolor=blue, urlcolor=blue]{hyperref}
\usepackage{txfonts}
\usepackage{lipsum}
\usepackage{subcaption}
\usepackage{gensymb}
\usepackage[flushleft]{threeparttable}
\usepackage{arydshln}

\begin{document}

\title{Estimating differential pistons for the Extremely Large Telescope using focal plane imaging and a residual network}


\author{P. Janin-Potiron\inst{1}\fnmsep\thanks{Corresponding author: \href{mailto:pierre.janin-potiron@lam.fr}{Pierre.Janin-Potiron@lam.fr}}
    \and
    M. Gray\inst{1}
    \and
    B. Neichel\inst{1}
    \and
    M. Dumont\inst{2,3,1}
    \and
    J.-F. Sauvage\inst{2,1}
    \and
    C. T. Heritier\inst{2,1}
    \and
    P. Jouve\inst{4,1}
    \and
    R. Fetick\inst{2,1}
    \and
    T. Fusco\inst{5,1}
}

\institute{Aix Marseille Univ, CNRS, CNES, LAM, Marseille, France
    \and
    DOTA, ONERA, 13330 Salon de Provence, France
    \and
    Faculdade de Engenharia da Universidade do Porto, Rua Dr. Roberto Frias, s/n, 4200-465 Porto, Portugal
    \and
    Space ODT - Optical Deblurring Technologies, Rua A. C. Monteiro, 65, 4050-014 Porto - Portugal
    \and
    DOTA, ONERA, Université Paris Saclay, 91123 Palaiseau, France
}

\titlerunning{ }
\authorrunning{Pierre Janin-Potiron et al.}

\date{}

\abstract
{As the Extremely Large Telescope (ELT) approaches operational status, optimising its imaging performance is critical. A differential piston, arising from either the adaptive optics (AO) control loop, thermomechanical effects, or other sources, significantly degrades the image quality and is detrimental to the telescope's overall performance.}
{In a numerical simulation set-up, we propose a method for estimating the differential piston between the petals of the ELT’s M4 mirror using images from a $2\times2$ Shack-Hartmann wavefront sensor (SH-WFS), commonly used in the ELT's tomographic AO mode. We aim to identify the limitations of this approach by evaluating its sensitivity to various observing conditions and sources of noise.}
{Using a deep learning model based on a ResNet architecture, we trained a neural network (NN) on simulated datasets to estimate the differential piston. We assessed the robustness of the method under various conditions, including variations in Strehl ratio, polychromaticity, and detector noise. The performance was quantified using the root mean square error (RMSE) of the estimated differential piston aberration.}
{This method demonstrates the ability to extract differential piston information from $2\times2$ SH-WFS images. Temporal averaging of frames makes the differential piston signal emerge from the turbulence-induced speckle field and leads to a significant improvement in the RMSE calculation. As expected, better seeing conditions result in improved accuracy. Polychromaticity only degrades the performance by less than $5\%,$ compared to the monochromatic case. In a realistic scenario, detector noise is not a limiting factor, as the primary limitation rather arises from the need for sufficient speckle averaging. The network was also shown to be applicable to input images other than the $2\times2$ SH-WFS data.}
{}


\maketitle

\section{Introduction}
\label{sec:introduction}

\begin{figure}[t]
    \centering
    \begin{subfigure}[b]{.49\linewidth}
        \centering
        {\includegraphics[width=.99\linewidth, trim=800 800 800 800, clip]{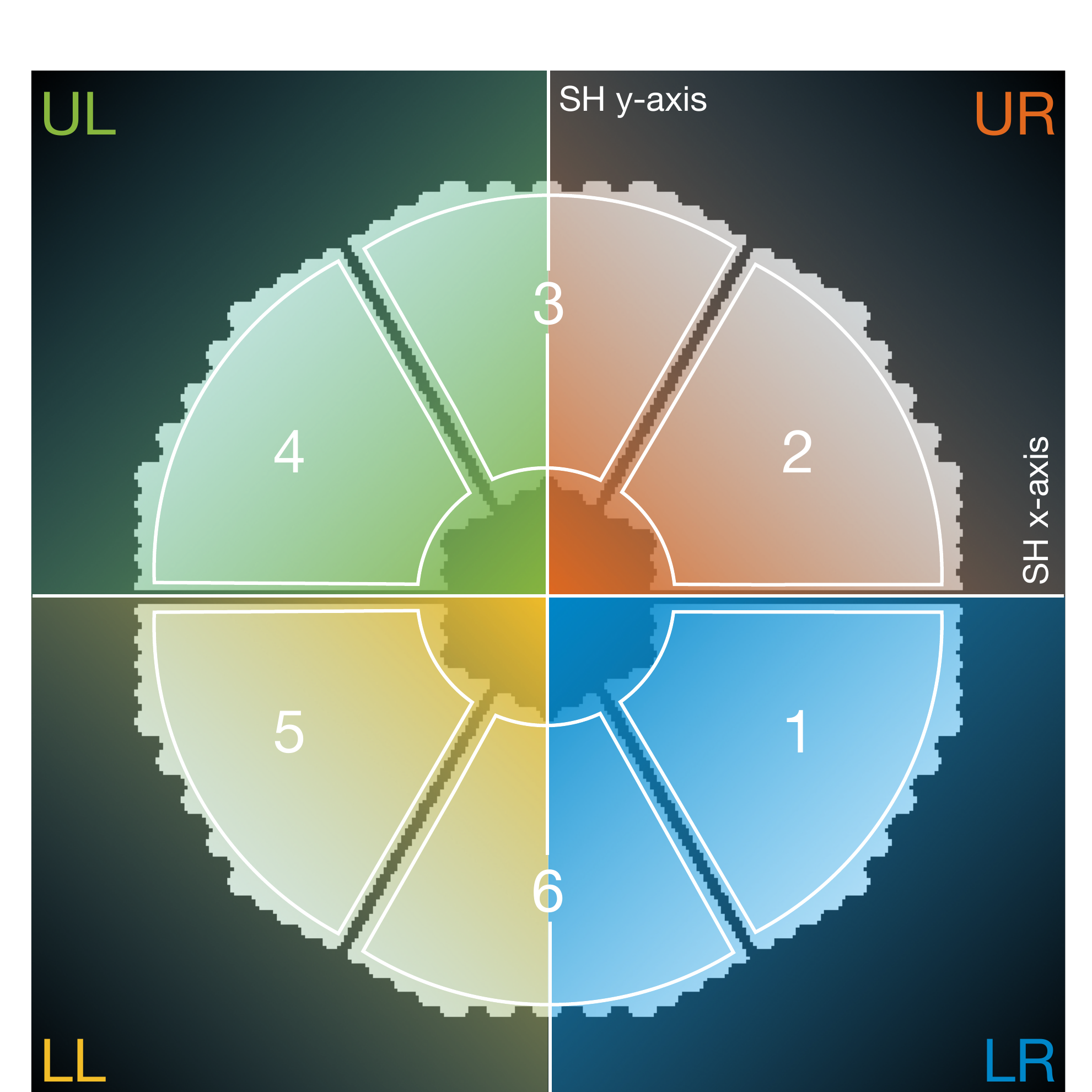}}
        \caption{}
        \label{fig:pupil:a}
    \end{subfigure}%
    \begin{subfigure}[b]{.49\linewidth}
        \centering
        {\includegraphics[width=.99\linewidth, trim=800 800 800 800, clip]{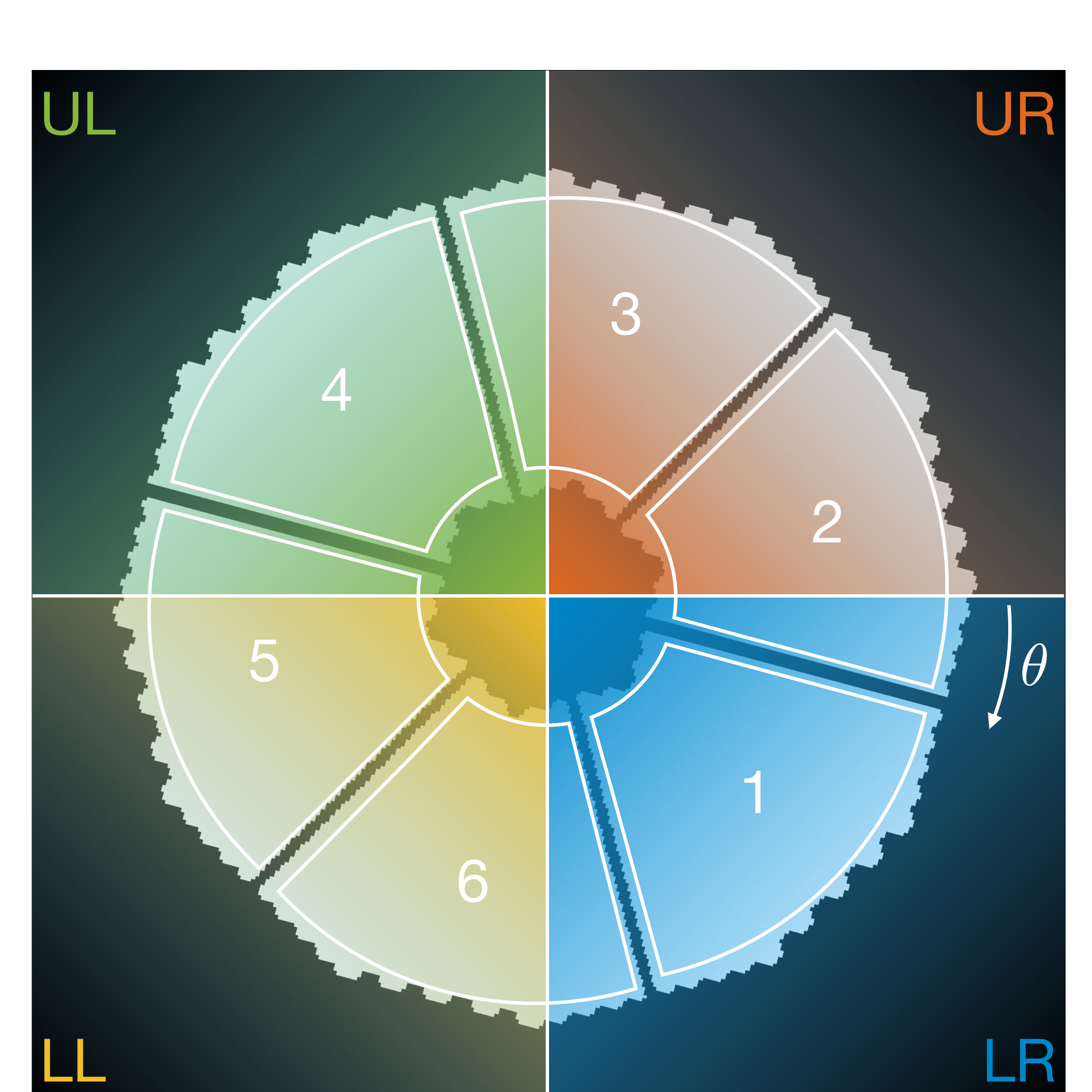}}
        \caption{}
        \label{fig:pupil:b}
    \end{subfigure}
    \begin{subfigure}[b]{.49\linewidth}
        \centering
        {\includegraphics[width=.47\linewidth]{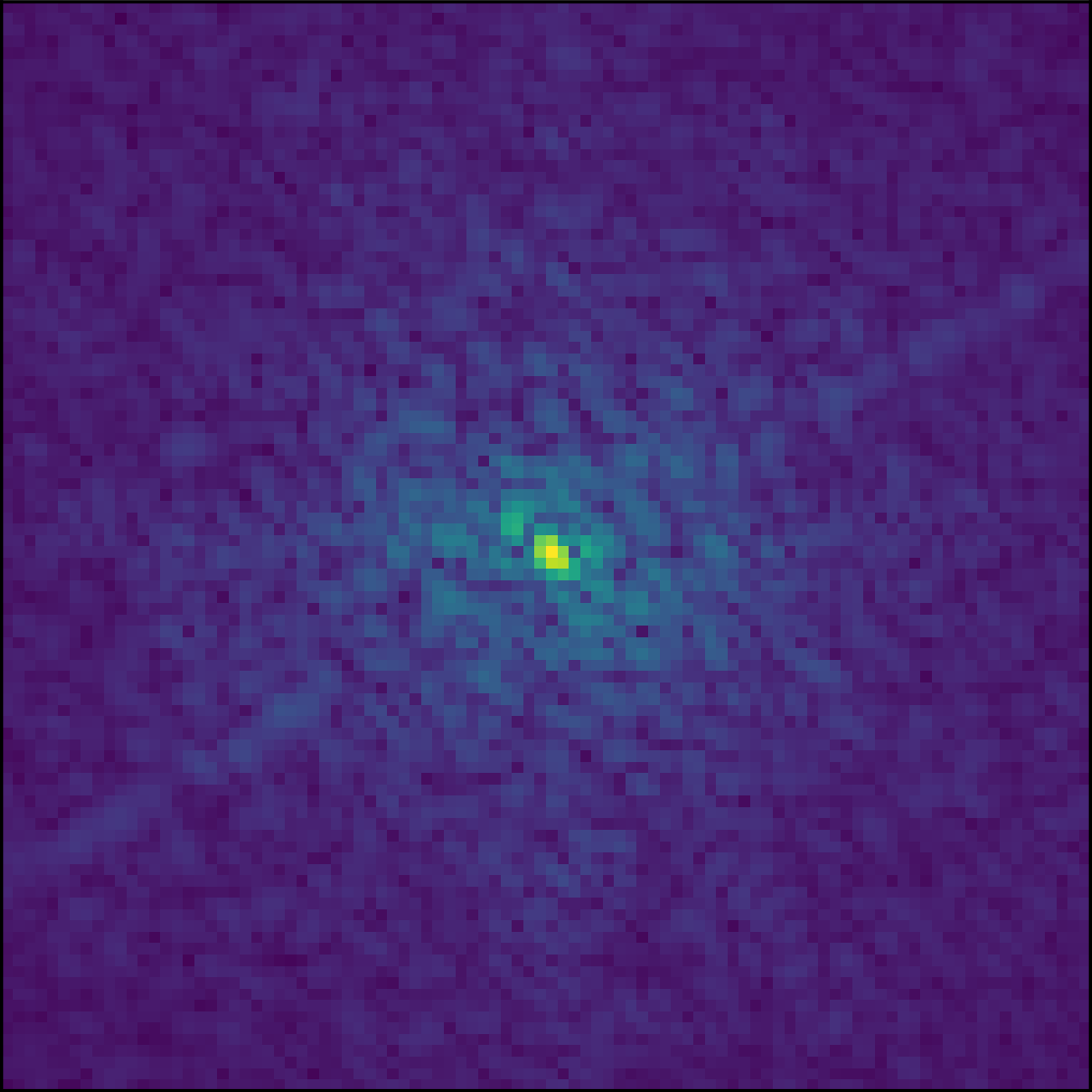}
            \includegraphics[width=.47\linewidth]{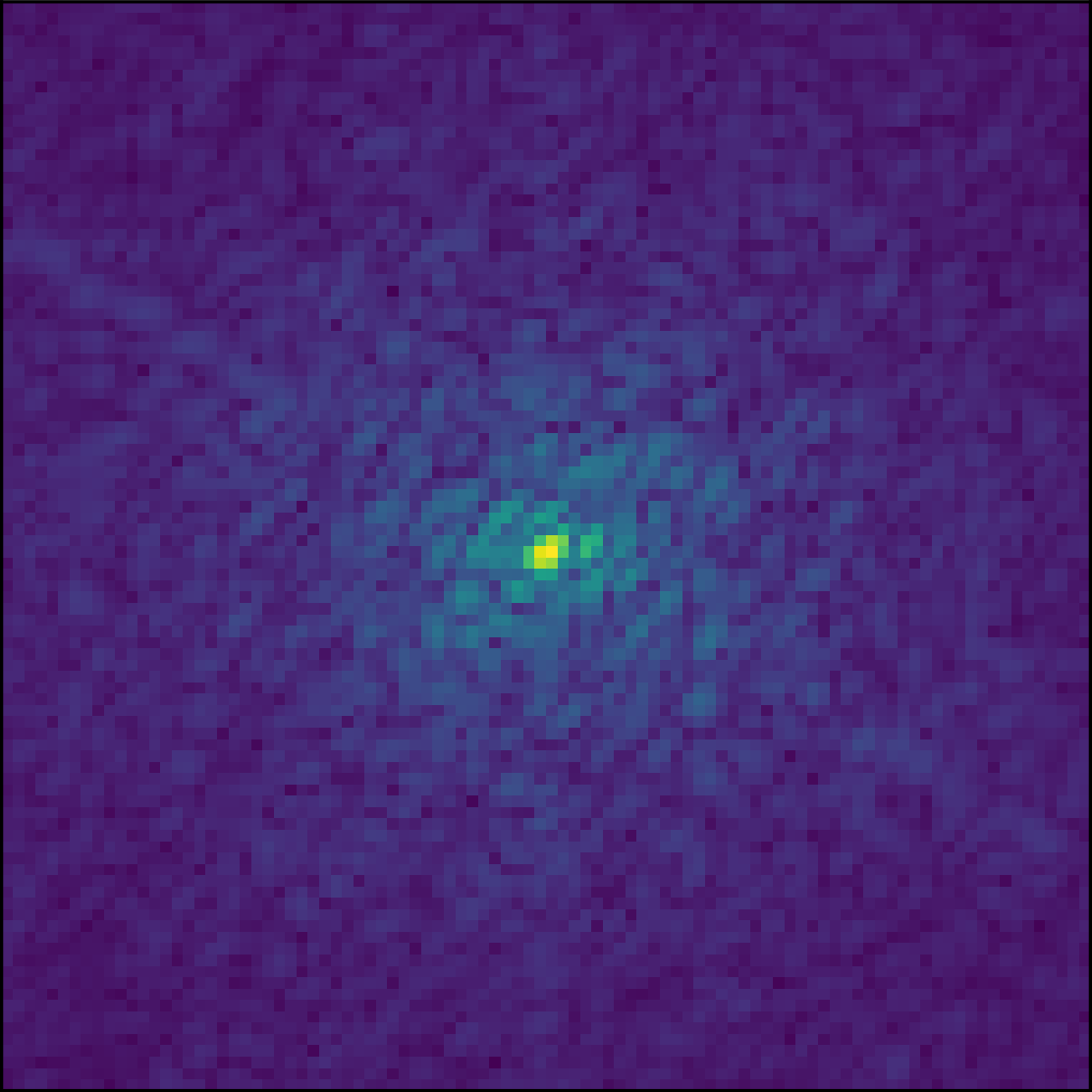}
            \includegraphics[width=.47\linewidth]{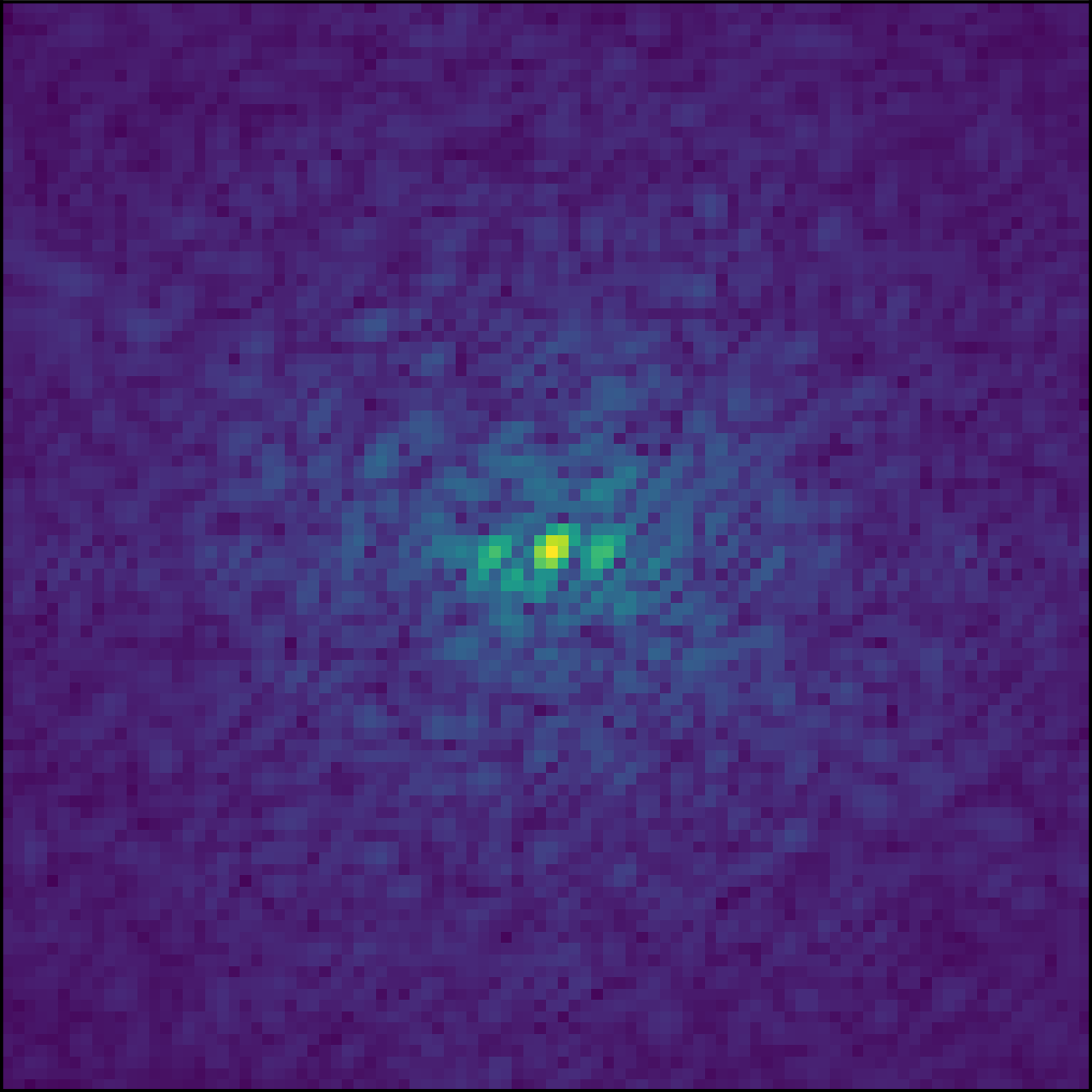}
            \includegraphics[width=.47\linewidth]{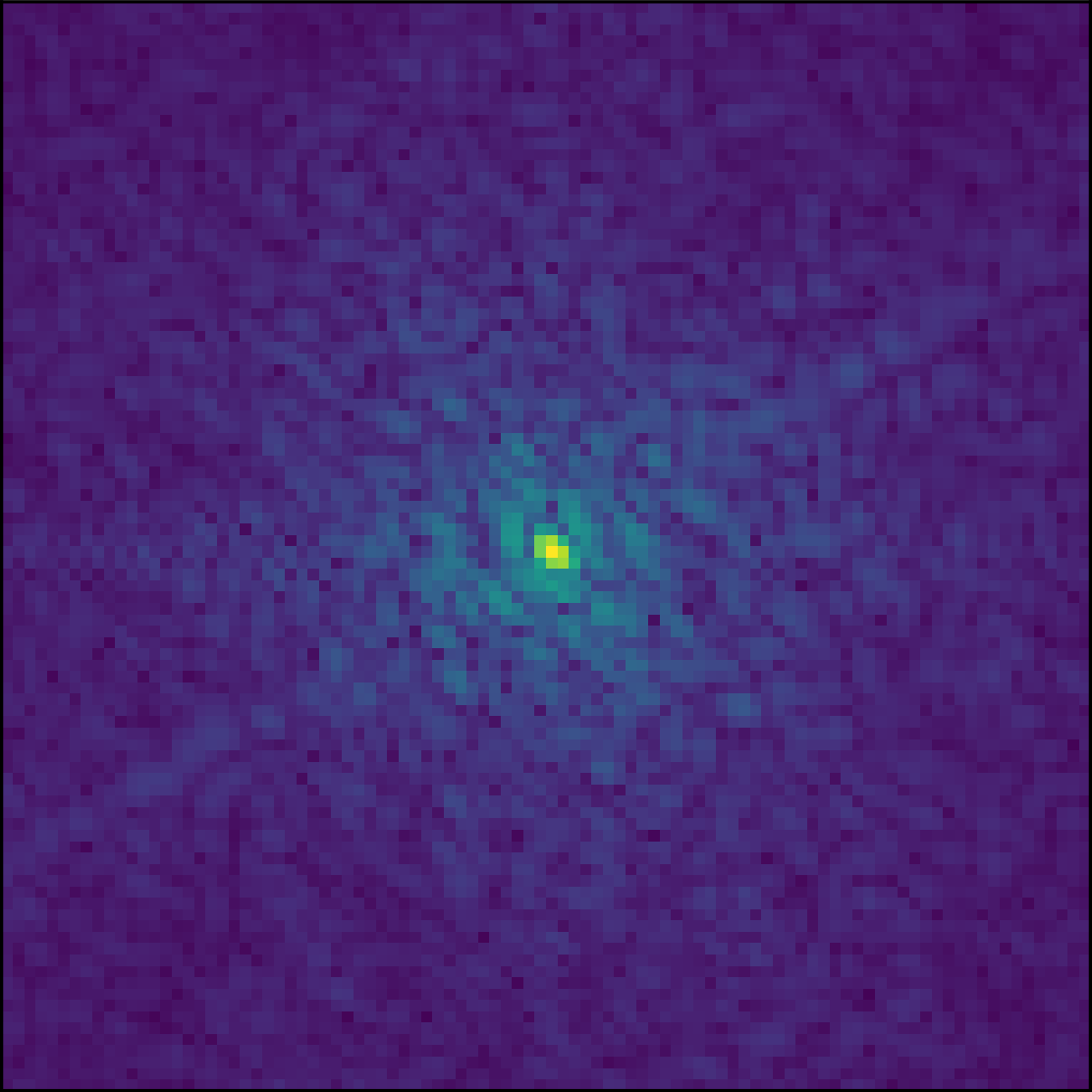}}
        \caption{}
        \label{fig:pupil:c}
    \end{subfigure}
    \begin{subfigure}[b]{.49\linewidth}
        \centering
        {\includegraphics[width=.47\linewidth]{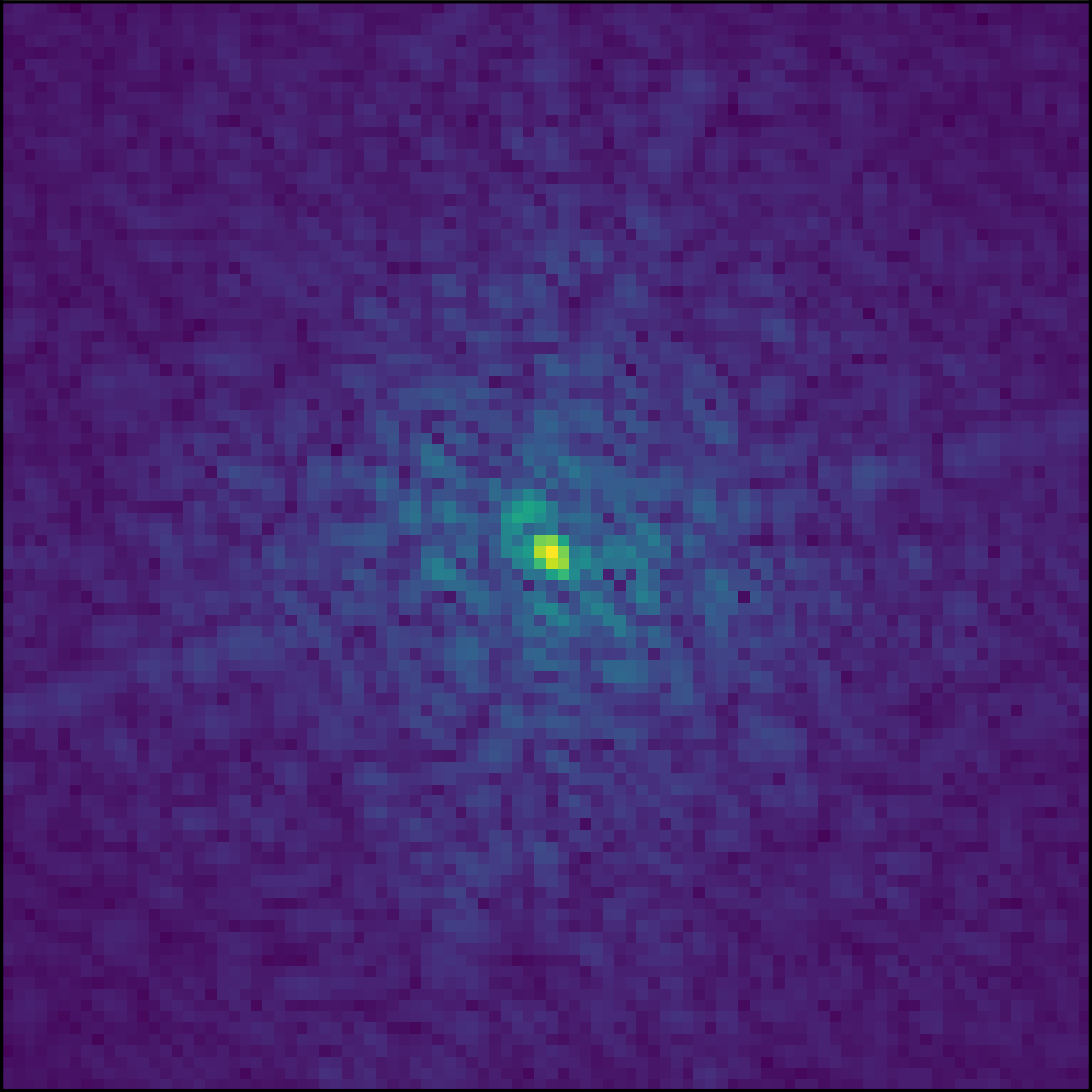}
            \includegraphics[width=.47\linewidth]{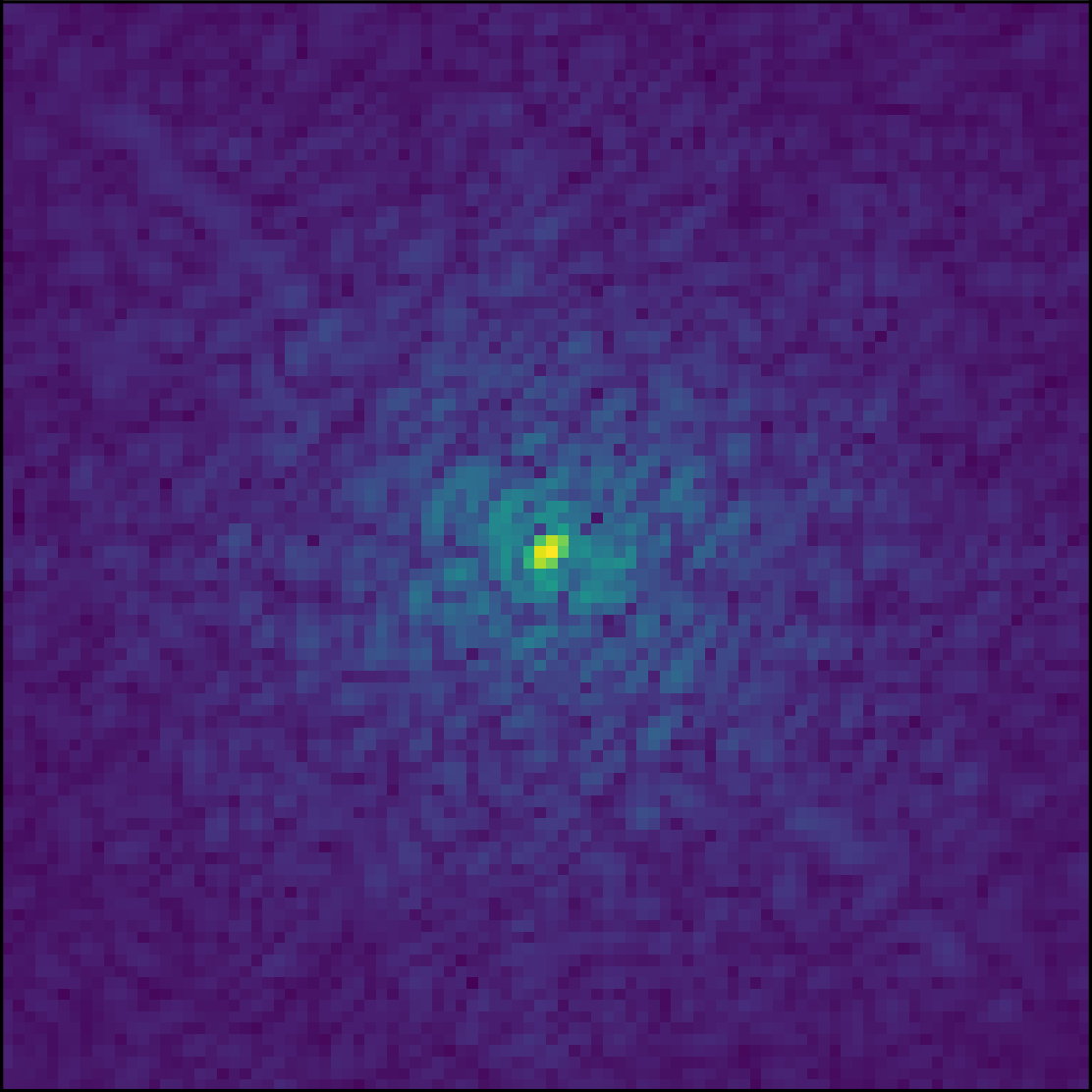}
            \includegraphics[width=.47\linewidth]{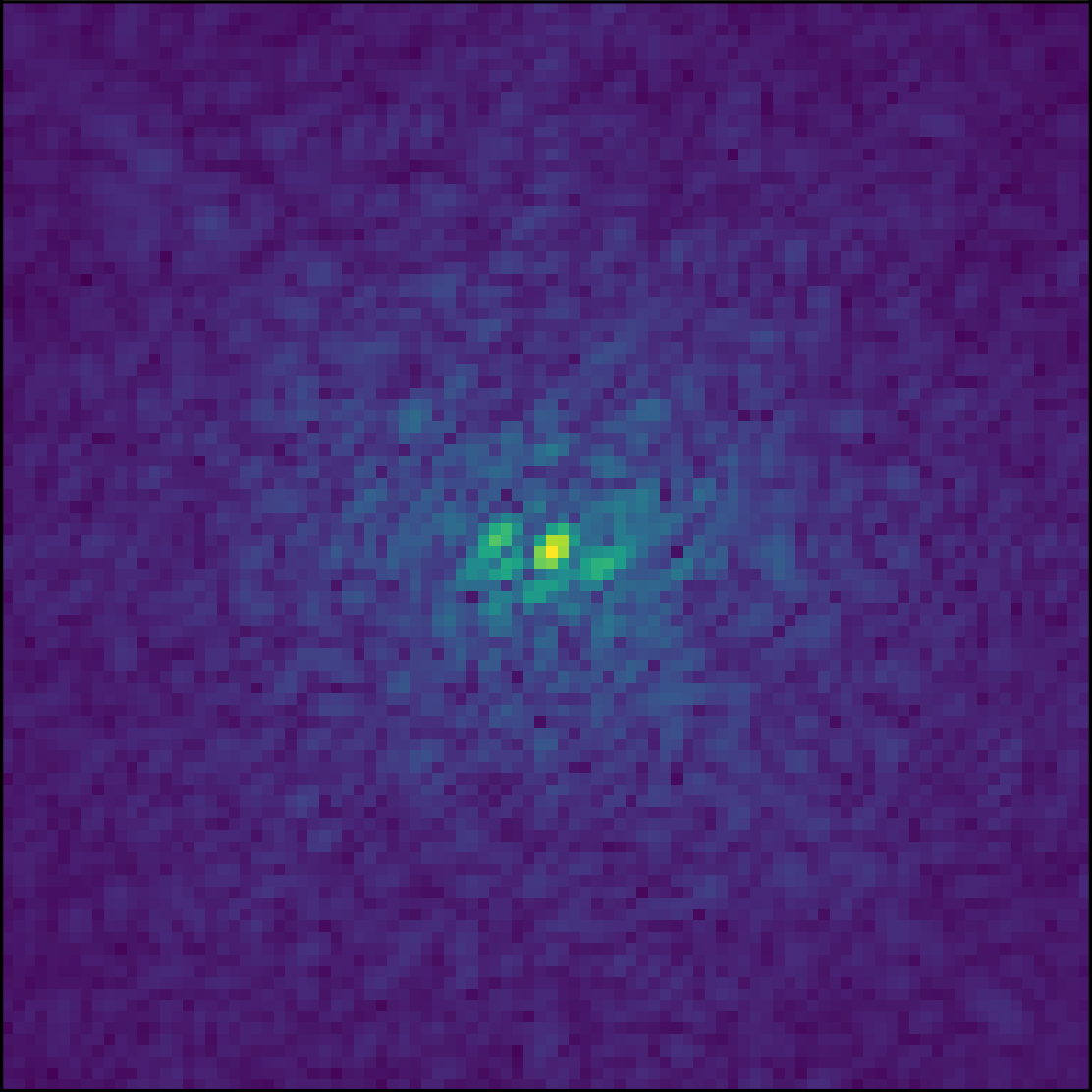}
            \includegraphics[width=.47\linewidth]{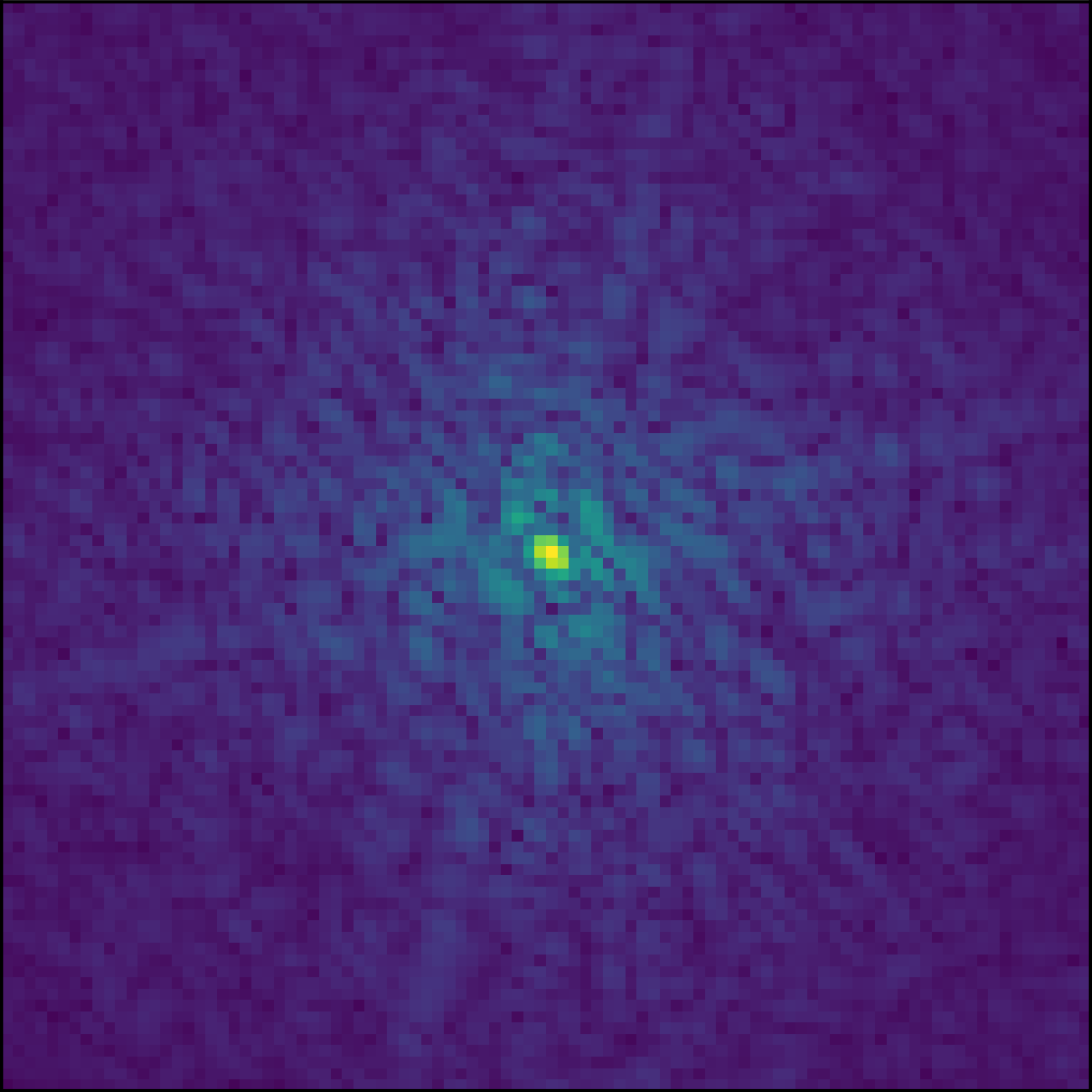}}
        \caption{}
        \label{fig:pupil:d}
    \end{subfigure}
    \caption{
        \label{fig:pupil}
        (a) Initial pupil configuration. The petals are numbered using the ESO convention.
        (b) Pupil configuration for a pupil rotation angle $\theta = 15\degree$.
        (c) and (d) Respective resulting sub-PSFs as measured by the $2\times2$ SH-WFS with atmospheric turbulence residuals.}
\end{figure}

The Extremely Large Telescope (ELT) is soon to become the largest ground-based optical telescope in operation, delivering unprecedented angular resolution that vastly surpasses the capacity of current facilities. With its 39 meter primary mirror, the ELT will allow astronomers to observe the Universe from the ground in unprecedented detail, achieving a resolution of 12 milliarcseconds in the K band, surpassing that of current VLT Unit Telescopes in the visible spectrum. Crucially, its performance will be enhanced by state-of-the-art adaptive optics (AO) systems, which will correct for atmospheric turbulence and allow the ELT to operate close to its diffraction limit. By combining vast light-gathering power with sharp image quality, the ELT is soon to become a transformative instrument for astronomy. It will address fundamental questions ranging from the search for extraterrestrial life and the study of the early Universe the nature of dark matter and dark energy, even along with detailed investigations of our own Solar System.

First generation ELT's intruments such HARMONI~\citep{Thatte:21}, MICADO~\citep{Davies:21}, and METIS~\citep{Brandl:21} will integrate AO systems~\citep{Bond:22,Heritier:24,Clenet:22,Feldt:24} relying on a pyramid wavefront sensor~\citep[PyWFS,][]{Ragazzoni:96}.
Despite its strengths, the modulated PyWFS exhibits intrinsically poor sensitivity to certain wavefront modes, including the so-called petaling effect (i.e. differential piston between the mirror segments of the ELT’s M4), which was initally identified as a limiting factor for optical performance by \citet{Schwartz:17}.
This manifestation of the effect arises from the pupil obscuration introduced by the spiders supporting M4, which disrupts phase continuity across the pupil and renders the sensor insensitive to this type of mode.
Several solutions have been proposed to improve the robustness of the AO system to this aberration by modifying the control strategy of the deformable mirror (DM) within the AO loop. This includes different approaches such as coupling the actuators of the DM at the edges of the spiders~\citep[refered to as minioning;][]{Schwartz:17}, redefining the wavefront reconstructor~\citep{Hutterer:18}, or constraining the modal control basis with spatial continuity criterions~\citep{Bertrou-Cantou:2020}.

While this type of approach shows a good level of performance in terms of compensating for the lack of sensitivity of the PyWFS, it fails to provide any generalisations to other sources of differential piston, most notably the low-wind effect~\citep[LWE, ][]{Sauvage:15,Milli:18}. According to European Southern Observatory (ESO), bursts of differential piston caused by LWE may reach peak-to-valley (PTV) amplitudes of up to $\simeq800$~nm (ESO, priv. comm.), posing a significant threat to optical performance. Accurately measuring and correcting these aberrations is therefore essential to preserve wavefront integrity and fully exploit the ELT’s observational capabilities.
Therefore, sensors capable of directly measuring differential piston, whether induced by the AO loop or attributable to the low-wind effect, are essential to ensure a robust correction of this aberration.
Several solutions have already been proposed, either by incorporating additional hardware~\citep{NDiaye:18,Haffert:22,Lombardi:23,Levraud:24} or by exploiting information that is already available in existing sensors~\citep{Rossi:22,Dumont:24,Taheri:24}.
Our approach is aligned with this philosophy, aimed at directly estimating the differential piston from sensor data without relying on modal assumptions.

In the tomographic AO (LTAO) mode of the ELT, natural guide stars (NGSs) are used to sense low-order modes (typically tip, tilt, and focus). To that end, the most standard configuration~\citep[that will be used in particular at ELT for HARMONI and MICADO, ][]{Plantet:22} involves a dedicated $2\times2$ Shack-Hartmann wavefront sensor (SH-WFS), which produces four sub-point spread functions (sub-PSFs, visible in Figs.~\ref{fig:pupil:c} and \ref{fig:pupil:d}) corresponding to the four quadrants of the pupil. Although this type of wavefront sensor (WFS) was not originally designed to measure differential piston, the sub-PSFs it produces inherently contain information about phase discontinuities.
In this paper, we propose using it to estimate the differential piston mode between the petals of M4.

Given the non-linear and complex nature of this inverse problem, artificial intelligence (deep learning techniques in particular) emerges as a compelling solution.
Deep neural networks (NNs) have shown remarkable success in both pattern recognition and regression tasks across a wide range of fields, including wavefront sensing analysis~\citep{Guo:2019, OrbanDeXivry:2021, Quesnel:2022, Dumont:24}. Their ability to learn complex, non-linear mappings directly from raw data makes NNs particularly well suited for the task of differential piston sensing in turbulent and noisy conditions, where conventional methods can be difficult to fine-tune.
Among their advantages, the computational efficiency at inference time, adaptability to varying observing conditions, and ability to generalise across different input data types are particularly appealing. With this method, we do not aim to propose a better approach to differential piston measurement, but, rather, to introduce a viable alternative for its estimation.
Our data-driven strategy takes advantage of the strengths of artificial intelligence in handling complex tasks to directly estimate differential piston from sub-PSFs.
Finally, while a rigorous benchmark against existing techniques remains challenging due to the wide variety of configurations reported in the literature, previous studies using NN approaches~\citep{Guo:2019, DuBose:2020,Gray:22,Rossi:2022, Guo:2022} have shown the potential of such approaches to improve sensor performance compared to traditionnal methods.

In \citet{Janin:24}, we validated via numerical simulations that a relatively simple ResNet~\citep{He:15, He:16} architecture is capable of retrieving differential piston from sub-PSFs affected by residual atmospheric turbulence. Building on these results, the present study offers an in-depth simulation analysis of the method's performance under more realistic observing conditions.
After detailing the technical set-up and notations used throughout the paper in Sect.~\ref{sec:method} and describing the training procedure along with the generation of the datasets employed for both training and testing the NN, we present the results of our method under various conditions in Sect.~\ref{sec:results}. These include the dependency on averaging (Sect.~\ref{sec:results:temporal_averaging}), on polychromatism (Sect.~\ref{sec:results:chromatism}), and on detector noise (Sect.~\ref{sec:results:noise}), as well as the robustness to mismatches between training and inference conditions (Sect.~\ref{sec:results:crosstests}).
In Sect.~\ref{sec:applications}, we demonstrate that our network is capable of estimating differential piston from alternative image sources by training and predicting on LIFT-like~\citep{Plantet:13} images.
Finally, we conclude with a discussion of our findings and outline potential directions for future research in Sect.~\ref{sec:conclusions}.

\section{Method}
\label{sec:method}

In this section, we present the technical set-up and notations used throughout the paper, followed by a detailed description of the differential piston estimation method. We also discuss the training procedure and the generation of the dataset used for training and testing the NN.

\subsection{Technical set-up}
\label{sec:method:setup}

This section details the instrumental set-up and notations used throughout the paper.
As introduced in Sect.~\ref{sec:introduction}, our algorithm targets the estimation of differential piston values from a PSF acquired using a 2x2 Shack-Hartmann wavefront sensor (SH-WFS) operating on an off-axis natural guide star (NGS) and located downstream of a high-order loop correction ensured by the laser guide stars (LGS) and a M4+M5 correction. To illustrate the method's core principles, a simplified schematic of the optical set-up is presented in Fig.~\ref{fig:pupil}.

Figure~\ref{fig:pupil:a} presents the entrance pupil of the ELT in its nominal configuration, segmented into six subsections by the spiders supporting the secondary mirror. These subsections are referred to as petals and, following ESO notation, numbered from 1 to 6.
The SH-WFS axes, depicted as white straight lines, subdivide the entrance pupil into four subpupils. We designate these subpupils as UL (upper left), UR (upper right), LL (lower left), and LR (lower right) for clarity and simplicity.

\begin{table}[tb]
    \begin{threeparttable}
        \caption[]{List and main characteristics of the datasets used in our simulations.}
        \label{tab:simulation_parameters}
        \centering
        \begin{tabular*}{\linewidth}{l@{\extracolsep{\fill}}r}
            \hline
            \noalign{\smallskip}
            Pupil & ELT pupil\textsuperscript{\hyperref[note:simulation_parameters:a]{(a)}}\\
            & Monolithic primary with $50$~cm spiders\\
            & Central obstruction 13.5 m\\
            & Resolution: \texttt{238 px}\\
            & Rotation angle: \texttt{[0\degree,30\degree]}\\
            \hline
            \noalign{\smallskip}
            Piston & Uniform distribution on the 6 petals\\
            & $[-\pi + \epsilon, \,\pi - \epsilon]$ range ($\epsilon=\pi/50$, see Sect.~\ref{sec:method:pi_ambiguity})\\
            \hline
            \noalign{\smallskip}
            Turbulence & From \texttt{TIPTOP}\textsuperscript{\hyperref[note:simulation_parameters:b]{(b)}} PSD\\
            & Independent or FATMOSS (see Sect.~\ref{sec:results:temporal_averaging})\\
            & Three different Strehl ratios (see Table~\ref{tab:datasets})\\
            \hline
            \noalign{\smallskip}
            Sub-PSF & Monochromatic or polychromatic (H-band)\\
            & Resolution: \texttt{96 px}\\
            & Sampling: \texttt{2 px per $\mathtt{\lambda/D}$}\\
            \hline
        \end{tabular*}
        \begin{tablenotes}
            \item \textsuperscript{\label{note:simulation_parameters:a}(a)} from \texttt{P3} library \citep{Beltramo:20}, \textsuperscript{\label{note:simulation_parameters:b}(b)}\cite{neichel2021tiptop}
        \end{tablenotes}
    \end{threeparttable}
\end{table}

\begin{table*}
    \centering
    \begin{threeparttable}
        \caption[]{List of the main characteristics of the datasets used in our simulations.}
        \label{tab:datasets}
        \begin{tabular}{p{.2\linewidth} p{.25\linewidth} p{.25\linewidth} p{.2\linewidth}}
            \hline
            \noalign{\smallskip}
            ID
                                                                  & Strehl ratio+seeing

            (H-band)
                                                                  & Chromatism ($\lambda_0 = 1.65\mu m$)
                                                                  & Atmosphere                                                                                                                                                   \\
            \noalign{\smallskip}
            \hline
            \noalign{\smallskip}

            SP10 (Fig.~\ref{fig:uncorrelated})                    & $10\%\,/\,1.042\mathrm{\arcsec}$         & monochromatic ($\Delta\lambda=0$)        & independent                                                            \\

            SP25 (Fig.~\ref{fig:uncorrelated}, \ref{fig:noise})   & $25\%\,/\, 0.568\mathrm{\arcsec}$        & monochromatic ($\Delta\lambda=0$)        & independent                                                            \\

            SP25 FATMOSS (Fig.~\ref{fig:uncorrelated_vs_fatmoss}) & $25\%\,/\, 0.568\mathrm{\arcsec}$        & monochromatic ($\Delta\lambda=0$)        & frozen-flow + boiling\textsuperscript{\hyperref[note:datasets:a]{(a)}} \\

            SP25 PC 0.3 (Fig.~\ref{fig:polychromatic})            & $25\%\,/\, 0.568\mathrm{\arcsec}$        & polychromatic ($\Delta\lambda=0.3\mu m$) & independent                                                            \\

            SP25 PC 0.6 (Fig.~\ref{fig:polychromatic})            & $25\%\,/\, 0.568\mathrm{\arcsec}$        & polychromatic ($\Delta\lambda=0.6\mu m$) & independent                                                            \\

            SP50 (Fig.~\ref{fig:uncorrelated})                    & $50\%\,/\, 0.432\mathrm{\arcsec}$        & monochromatic ($\Delta\lambda=0$)        & independent                                                            \\

            SP20-30 (Fig.~\ref{fig:crosstest_SP_range})           & $20-30\%$                            N/A & monochromatic ($\Delta\lambda=0$)        & independent                                                            \\

            SP10-50 (Fig.~\ref{fig:crosstest_SP_range})           & $10-50\%$                            N/A & monochromatic ($\Delta\lambda=0$)        & independent                                                            \\

            \noalign{\smallskip}
            \hline
        \end{tabular}
        \begin{tablenotes}
            \item \textsuperscript{\label{note:datasets:a}(a)} from the \texttt{FATMOSS} library \citep{Kuznetsov:24}
        \end{tablenotes}
    \end{threeparttable}
\end{table*}

In addition, the telescope's tracking motion induces a relative rotation between the pupil and the SH-WFS axes, as illustrated in Fig.~\ref{fig:pupil:b}. We define this rotation angle, denoted by $\theta$, as the angle formed between the SH-WFS x-axis and the axis separating petals 1 and 2 of the pupil. In this specific case, for consistency with our previous work~\citep{Chauvet:23}, the rotation angles increase as they rotate in the clockwise direction.
Due to symmetry, analysing only subpupils UR and UL for a rotation angle $\theta \in [0\degree,30\degree]$ is sufficient to characterise the differential piston effects. This approach reduces the number of differential piston values to be considered to one for the UR sub-pupil ($\Delta p_{\{3,2\}}$) and two for the UL sub-pupil ($\Delta p_{\{4,3\}}$ and $\Delta p_{\{5,4\}}$).
In this paper, we focus exclusively on the UR subpupil. As shown in our previous work \citep{Janin:24}, the UR and UL subpupils yield similar results. Therefore, findings for the UR case are expected to generalise directly to both UR and UL subpupils.

Each subpupil generates a corresponding PSF, referred to as a sub-PSF, as shown in Figs.~\ref{fig:pupil:c} and \ref{fig:pupil:d},. The PSFs exhibit characteristic diffraction patterns that arise due to two primary factors: (1) the presence of the telescope's spider and (2) the square cropping of the entrance pupil. Figure~\ref{fig:pupil:d} illustrates that when the pupil is rotated relative to the SH-WFS axes, the PSFs of subpupils containing three petals (UL and LR in Fig.~\ref{fig:pupil:b}) exhibit more intricate diffraction patterns.

The six petals of the entrance pupil are susceptible to piston errors, which manifest as a uniform displacement of an entire petal along the optical axis. These errors can arise from various sources, including atmospheric turbulence, the AO system itself, and mechanical or thermal effects. While a global piston shift across the entire pupil does not degrade image quality, differential piston, also known as petaling, occurs when individual petals exhibit different piston errors. This differential piston significantly impacts the final image quality.
Here, we define the differential piston between petals $i$ and $j$ as the difference in their respective piston values $p_i$ and $p_j$, thereby yielding
\begin{equation}
    \label{eq:diff_piston}
    \Delta p_{\{j,i\}} = p_j - p_i ~\mathrm{with}~ i,j \in [1;6],~i\neq j \,.
\end{equation}

We note that in this paper, the differential piston naturally present in the turbulence residuals is not included in the piston budget. This is because the SH-WFS operates with an off-axis NGS, so the differential piston induced by turbulence differs from the on-axis one. It is only if some degree of correlation would exist between the on-axis and off-axis differential pistons that the network could implicitly learn to infer a fraction of the on-axis differential piston from the off-axis measurements. However, no such correlation has been established at this stage to our knowledge. As a result, our network is effectively trained to measure only the telescope-induced differential piston in presence of residual piston contributions from turbulence.

\subsection{Numerical approach}
\label{sec:method:numerical_approach}

We simulated the technical set-up described in Sect.~\ref{sec:method:setup} using the \texttt{aoSystems} tool from the \texttt{P3} library \citep{Beltramo:20}, along with a custom Python library developed in-house. Table~\ref{tab:simulation_parameters} summarises the main parameters described in the following paragraph. We modeled the ELT primary mirror with its six petals, each assigned a random piston value following the procedure detailed in Sect.~\ref{sec:method:piston_distribution}.

Atmospheric turbulence is simulated as a single ground-layer phase screen, generated from a power spectral density (PSD) obtained from the \texttt{TIPTOP} library \citep{neichel2021tiptop}. We considered three configurations, corresponding to different Strehl ratios, by varying the seeing conditions (see Table~\ref{tab:datasets}).
In all simulations, the phase screens are uncorrelated, ensuring that each image within a sequence is independent. We describe one exception in Sect.~\ref{sec:results:temporal_averaging:correlated}, where we assess the temporal robustness of our algorithm. For this specific case, we used a frozen flow plus boiling phase screen generated using the \texttt{FATMOSS} library \citep{Kuznetsov:24}, resulting in correlated images within the sequence.

To mimic the action of the $2\times2$ SH-WFS, the full pupil was subdivided into four distinct sub-pupils. The sub-PSF images were obtained by computing the squared modulus of the Fourier transform of each sub-pupil's complex amplitude. For that particular operation, we used the matrix Fourier transform method \citep[MFT, ][]{Soummer:07}, which enables the efficient generation of $\mathtt{96}\times\mathtt{96}\,\mathrm{px^2}$ sub-PSFs (sampled at the Shanon limit) from a $\mathtt{238}\times\mathtt{238}\,\mathrm{px^2}$ ELT pupil.

In Section \ref{sec:results:chromatism}, we evaluate the system's performance under polychromatic conditions. Polychromatic images are simulated by summing multiple monochromatic images, each acquired at a distinct wavelength within the optical bandpass of the filter under consideration.
To accurately mimic polychromatism, we express a criterion for wavelength sampling as follows: the difference between the outermost diffraction orders contained in the image of the two shortest wavelengths must not exceed half a pixel. This leads to the following condition,
\begin{equation}
    n_{bins} \geq \left(\frac{\Delta\lambda}{\lambda_0-\Delta\lambda/2}\right) \times N_{array}\,,
\end{equation}
where $n_{bins}$ is the number of bins taken to sample the bandwidth, $N_{array}$ is the size (in pixels) of the image array, $\lambda_0$ is the central wavelength, and $\Delta\lambda$ is the bandwidth. The central wavelength is consistently set to $\lambda_0 = 1.65\,\mu\mathrm{m}$ throughout the paper, while the bandwidth $\Delta\lambda$ varies depending on the case considered (see Table~\ref{tab:datasets}).
In pratice, we have, for example, $n_{bins}\,\big\rvert\,_{\Delta\lambda = 0.3\mu m} > 19$.

\subsection{Piston distribution}
\label{sec:method:piston_distribution}

To proceed with our simulations, a specific statistical model for differential piston must be assumed. Although the true distribution on the ELT remains unquantified at present, we expect that ongoing end-to-end studies by ESO will soon provide crucial insights on its characteristics.
Given this actual uncertainty, we arbitrarily chose to model the individual piston values as following a uniform distribution. Thus, we obtained\begin{equation}
    \label{eq:piston_distribution}
    p_i \sim \mathcal{U}(-\pi/2, \pi/2) ~\mathrm{with}~ i \in [1;6]\,.
\end{equation}
This choice has the direct consequence of yielding a triangular distribution for the differential piston, which is the quantity of interest for our analysis. Thus, we obtained
\begin{equation}
    \label{eq:piston_distribution}
    \Delta p_{\{j,i\}} \sim \mathcal{T}(-\pi, \pi) ~\mathrm{with}~ i,j \in [1;6]\,,~i\neq j\,.
\end{equation}
Annex tests performed with a uniform distribution showed no significant differences in the final results.

\subsection{$2\pi$ ambiguity}
\label{sec:method:pi_ambiguity}

\begin{figure}
    \centering
    \includegraphics[width=\linewidth]{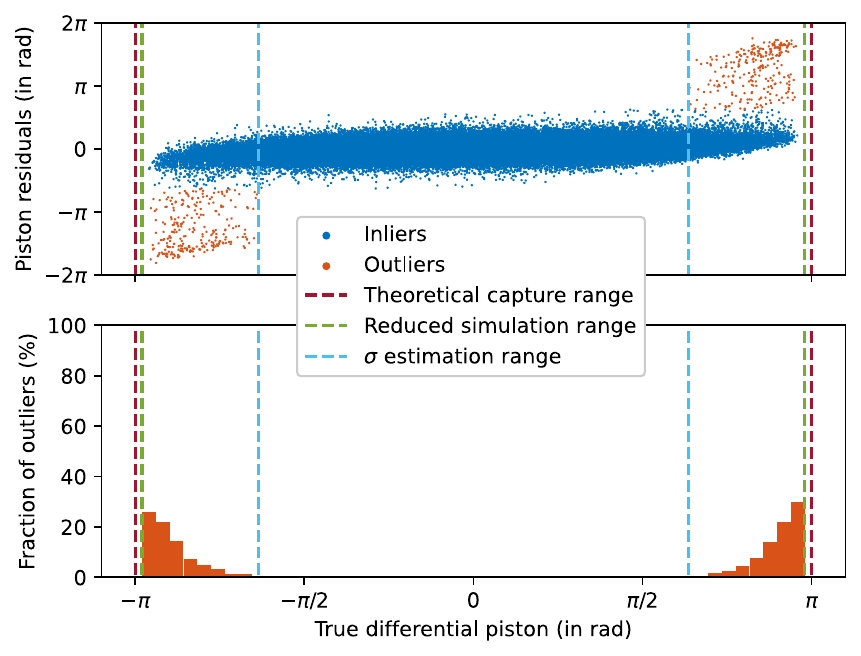}
    \caption{Outliers and range definition. Top: Residual differential piston as a function of the true differential piston for single frame SP10 configuration. The blue points represents the inliers, while the red points are the outliers. Bottom: Histogram of the outliers distribution.}
    \label{fig:scatter}
\end{figure}

The 2$\pi$ ambiguity, arising from the periodic nature of light waves, represents a fundamental limitation inherent in phase analysers \citep[e.g.][]{Pinna:06,Surdej:11,Janin:16, Janin:17}. This ambiguity restricts phase measurements to the interval $[-\pi,\pi]$ radians, commonly referred to as the capture range. While multi-wavelength methods exist to extend this capture range \citep{Chanan:98, Pinna:06, Surdej:11, Vigan:11, Martinez:16,Haffert:22}, they were not considered in this study. We assumed that these methods are equally applicable to our algorithm and their discussion beyond the scope of the present paper.

Edge effects, directly attributable to the 2$\pi$ ambiguity, manifest during differential piston estimation, leading to erroneous estimations near the theoretical limits of $-\pi$ and $+\pi$. These inaccurate estimations are termed outliers and are represented as red dots on Fig.~\ref{fig:scatter}. To mitigate these effects, we opted to train our network with a reduced capture range compared to the theoretical one. This adjusted interval spans $[-\pi+\epsilon, \pi-\epsilon]$, where $\epsilon=\pi/50$. Despite this precaution, outliers were still present in the piston values predicted by the NN. Consequently, we addressed them by manually clipping values exceeding $5\sigma$ (where $\sigma$ is computed over the range delimited by the light blue vertical lines in Fig.~\ref{fig:scatter}, corresponding to a $\pm2\,\text{rad}$ interval) a method visually confirmed to be effective in Fig.~\ref{fig:scatter}.
We denote the inliers set as $\mathcal{I}$ and the outliers set as $\mathcal{O}$ in the following of the paper.

According to preliminary estimates provided by ESO, the expected differential piston originating from LWE for the ELT reaches approximately 800 nm PTV (ESO, priv. comm.). This value, while substantial, remains within the theoretical capture range of our algorithm in the H band, suggesting that the proposed method should operate effectively under the anticipated on-sky conditions.
However, as discussed in Sect.~\ref{sec:introduction}, the telescope is expected to experience petaling bursts originating from the AO loop itself, which are not addressed by the previously discussed methods. These short events, assumed to be as fast as a few tens of ms, may compound with the LWE, potentially pushing the aberration beyond the capture range. Possible solutions to this issue are discussed in Sect.~\ref{sec:conclusions}.

\subsection{Evaluation metrics}
\label{sec:method:metrics}

The optical performance of our method is evaluated using the root mean squared error (RMSE), computed exclusively on the inlier data and defined as,
\begin{equation}
    \label{eq:rmse}
    \mathrm{RMSE} = \sqrt{\frac{1}{N}\sum_{k=1}^{N}\left(\widehat{\Delta p^k_{\{3,2\}}} - \Delta p^k_{\{3,2\}}\right)^2},\,p_{\{3,2\}}^k \in \mathcal{I},
\end{equation}
where $\widehat{\Delta p^k_{\{3,2\}}}$ is the estimated value of $\Delta p^k_{\{3,2\}}$.
Outliers are systematically excluded from this calculation using the method detailed in Sect.~\ref{sec:method:pi_ambiguity}.
As stated in Sect.~\ref{sec:method:piston_distribution}, we chose a uniform distribution for the pistons. Consequently, the baseline RMSE of our simulation is
\begin{equation}
    \label{eq:rmse}
    \mathrm{RMSE}\,\big\rvert\,_{\mathcal{U}} = \frac{\pi \sqrt{6}}{6} (\mathrm{rad}) = \frac{\lambda \sqrt{6}}{12} (\mathrm{m})\,,
\end{equation}
for $\lambda = 1.65\mu m$, this gives $\mathrm{RMSE}\,\big\rvert\,_{\mathcal{U}} \simeq 337\,\mathrm{nm}$.

\section{Deep learning}

\subsection{Neural network architecture}
\label{sec:method:nn_architecure}

The NN is based on a residual convolutional architecture designed for scalar regression (i.e. retrieve a scalar differential piston value).
The input of the NN consists of single-channel sub-PSFs of that are $96\times96\times1$ in size,
which are normalised so that their integral value is 1. In noiseless situations
(Sect.~\ref{sec:results:temporal_averaging}, \ref{sec:results:chromatism},
\ref{sec:results:crosstests:average}, \ref{sec:results:crosstests:strehl})
a min-max normalisation must be applied to ensure numerical stability and facilitates
convergence during optimisation.

\begin{figure}
    \centering
    {\includegraphics[page=2,height=0.38\linewidth]{figures/residual_layer.pdf}
        \includegraphics[page=1,height=0.38\linewidth]{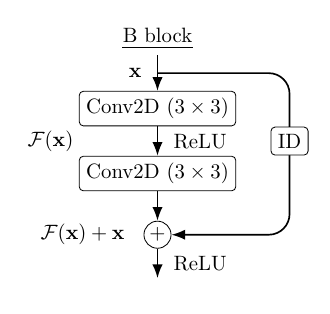}}
    \caption{
        \label{fig:blocks}
        Residual blocks used in the ResNet. Left: A-block type, with the skip layer being a $1\times1$ convolution. Right: B-block type, with the skip layer being the identity.}
\end{figure}

\begin{figure}
    \centering
    \includegraphics[page=3,width=.75\linewidth]{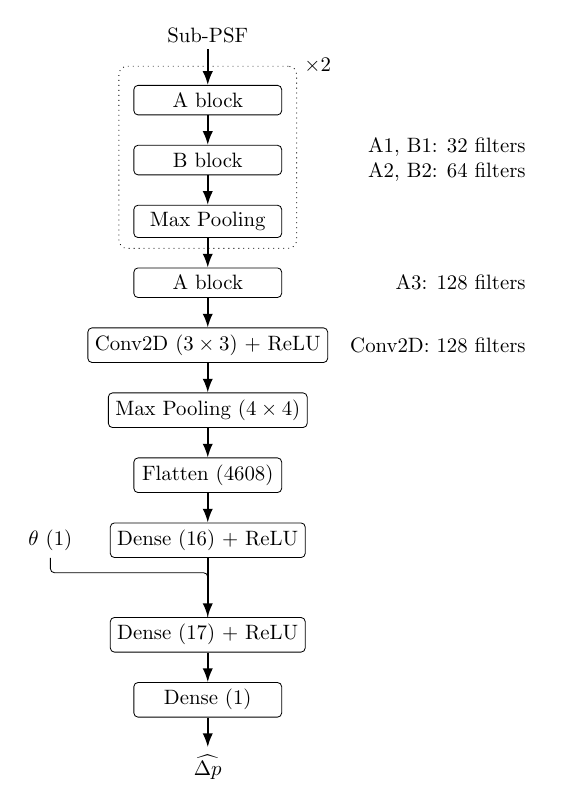}
    \caption{
        \label{fig:network}
        ResNet architecture. The number of filters is shown to the right of each convolutional layer.
    }
\end{figure}

The NN includes two consecutive stages composed of residual blocks of types A and B (see Fig.~\ref{fig:blocks}). Each block consists of two convolutional layers with ReLU activations and a shortcut connection that preserves the number of output filters, either an identity mapping for type A or a $1\times1$ convolution for type B. A final ReLU activation is applied after each residual summation.
Max-pooling layers with $2\times2$ kernels and stride 2 follow selected blocks to progressively reduce spatial resolution.
A final convolutional layer with ReLU activation is applied after these two residual stages, followed by a $4\times4$ max-pooling operation. At this stage, a scalar input representing the known pupil rotation angle, $\theta$, is concatenated with the output of the dense layer. The resulting vector is then passed through two additional dense layers with an intermediate ReLU activation, yielding a single scalar output corresponding to the estimated differential piston.
In all NN architectures, the convolutional layers use a kernel size of $3\times3$ and a stride of 1, except for the skip connections in the B block, which use $1\times1$ convolutions. The number of filters increases progressively from 32 to 128 across the network, as detailed in Fig.~\ref{fig:network}.
Batch normalisation is not considered in our network design.
A diagram of the full architecture is provided in Fig.~\ref{fig:network}.

\subsection{Training procedure}
\label{sec:method:training}

The NN is trained to predict differential piston values from simulated sub-PSFs in a supervised regression architecture using the \texttt{DEEPLOOP} library~\citep{Gray:22}.
We used datasets of $360\,000$ simulated sub-PSFs, uniformly distributed over a range of input pupil rotation angles, resulting in exactly $12\,000$ sub-PSFs per $1\degree$ interval. The differential piston values follow a uniform distribution, as described in Sect.~\ref{sec:method:piston_distribution}.
We selected $300\,000$ images for training and keep an additional $60\,000$ for validation. Ground-truth piston values are directly derived from the simulation inputs.
The model was trained to minimise the Huber loss (selected for its robustness in handling outliers), defined as
\begin{equation}
    \label{eq:loss}
    \mathcal{L}_\delta(\Delta \phi, \widehat{\Delta \phi}) =
    \begin{cases}
        \frac{1}{2}\left(\Delta \phi - \widehat{\Delta \phi}\right)^2                               & \text{if } \left|\Delta \phi - \widehat{\Delta \phi}\right| \leq \delta, \\
        \delta \left( \left|\Delta \phi - \widehat{\Delta \phi} \right| - \frac{1}{2}\delta \right) & \text{otherwise},
    \end{cases}
\end{equation}
where $\Delta \phi = 2\pi \Delta p / \lambda$ is the ground-truth differential phase value, $\widehat{\Delta \phi}$ is its predicted value, and $\delta$ is a threshold that we set to $\delta = 1\,\text{rad}$.
We use the Adam optimiser \citep{Kingma:14} with a batch size of 128 and an adaptive learning rate with optimised parameters selected via a grid search.
The training was conducted on 150 epochs and we retain the model weights from the epoch yielding the lowest validation loss (Eq.~\ref{eq:loss}).
No regularisation or data augmentation is applied, as the dataset already contains substantial variability in turbulence configurations and pupil rotation angles.

\section{Results}
\label{sec:results}

Our initial study~\citep{Chauvet:23} validated the capability of the NN to measure differential piston from images affected by simplified atmospheric turbulence. Building on this work, \citet{Janin:24} further confirmed the efficacy of the system under more realistic conditions, including turbulence residuals from AO. This second study also showed that averaging multiple images significantly enhances the performance of differential piston estimation.

In this section, we investigate the robustness of the NN with respect to key variables that can affect its performance. In particular, we assess the dependence on the number of averaged images (Sect.~\ref{sec:results:temporal_averaging}), the influence of polychromaticity (Sect.~\ref{sec:results:chromatism}), and the impact of noise (Sect.~\ref{sec:results:noise}). In addition, Sect.~\ref{sec:results:crosstests} evaluates the network’s intrinsic robustness, where the system is trained under one set of conditions and tested under different ones. The RMSE results presented in this section were computed on dedicated test sets comprising a total of 90\,000 images, divided into 3\,000 images per rotation angle range of $1\degree$.

\subsection{Temporal averaging}
\label{sec:results:temporal_averaging}

\begin{figure}[t]
    \centering
    \includegraphics[height=.9\linewidth]{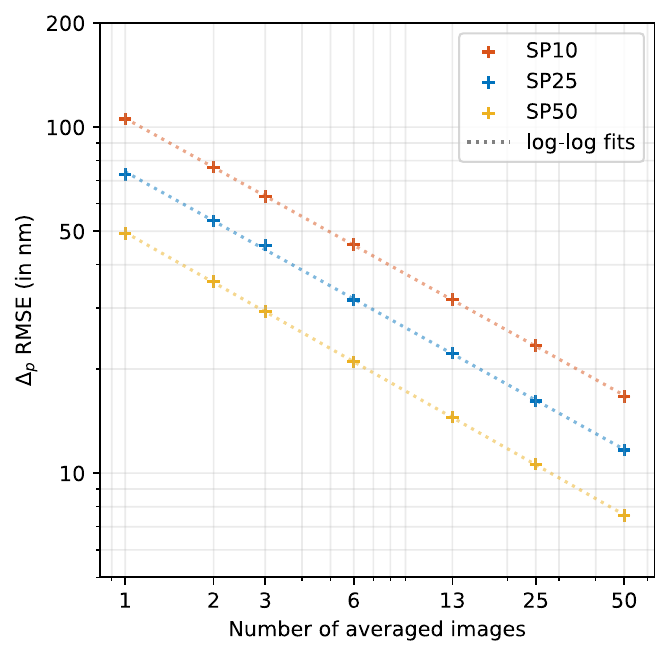}
    \caption{Influence of the number of averaged frames on the differential piston RMSE.
        Differential piston RMSE versus number of averaged images. Results are presented for the three turbulence configurations SP10, SP25, and SP50 listed in Table~\ref{tab:datasets} in a monochromatic uncorrelated regime.}
    \label{fig:uncorrelated}
\end{figure}

\begin{figure}[t]
    \centering
    \includegraphics[height=.9\linewidth]{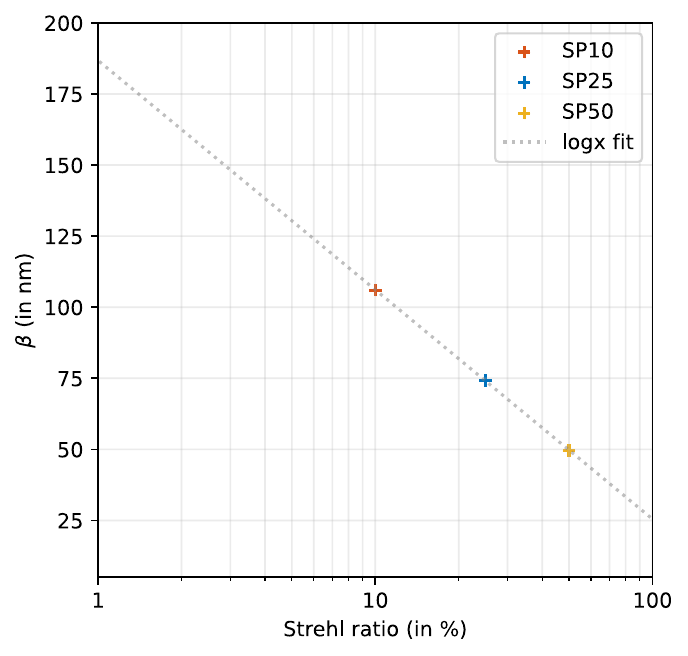}
    \caption{Influence of residual PSF Strehl ratio on differential piston RMSE.
        $\beta$ parameter (Eq.~\ref{eq:RMSE_n}) versus Strehl ratio for the three turbulence configurations SP10, SP25, and SP50 listed in Table~\ref{tab:datasets} in a monochromatic uncorrelated regime.}
    \label{fig:uncorrelated_strehl}
\end{figure}

As shown in \citet{Janin:24}, the accuracy of differential piston estimation can be significantly improved by averaging multiple images. In this section, we investigate the impact of temporal averaging on differential piston estimation under three different atmospheric conditions.

We distinguish between two cases of temporal averaging of sub-PSF images. In the first case, described in Sect.~\ref{sec:results:temporal_averaging:independent}, each frame is generated using an independent atmospheric phase screen. This approach assumes complete decorrelation of the turbulence between exposures, enabling faster simulation and the exploration of a wider range of configurations. This is the configuration used for all tests presented in this paper (apart from the exception described in Sect.~\ref{sec:results:temporal_averaging:correlated}).

In contrast, the second case, presented in Sect.~\ref{sec:results:temporal_averaging:correlated}, uses a temporally correlated sequence of phase screens. This simulates realistic atmospheric evolution, modeled as frozen flow with additional boiling using the \texttt{FATMOSS} library\citep{Kuznetsov:24}. This approach highlights the temporal behaviour of our method and provides characteristic timescales for the evolution of the RMSE.

\subsubsection{Independent phasescreens}
\label{sec:results:temporal_averaging:independent}

To begin evaluating the dependence on averaging, we analysed the effect of averaging an increasing number of independent sub-PSF images on the accuracy of differential piston estimation, for the three conditions (SP10, SP25, and SP50) presented in Table~\ref{tab:datasets}.

For each configuration, we computed the RMSE as a function of the number of averaged frames. The results (shown in log-log scale in Fig.~\ref{fig:uncorrelated}) demonstrate a clear improvement in RMSE with increasing numbers of averaged frames. The curves follow a power-law decay, which can be expressed as
\begin{equation}
    \label{eq:RMSE_n}
    \mathrm{RMSE}(n_{uncorr})= \beta \cdot n_{uncorr}^\alpha \,,
\end{equation}
where $n_{uncorr}$ is the number of averaged images, $\alpha$ is the power-law exponent, and $\beta$ is the RMSE for a single frame (i.e. $\beta = \mathrm{RMSE}(n_{uncorr}=1)$). For ease of comparison, Table~\ref{tab:RMSE} summarises the fitted values of $\alpha$ and $\beta$.

The observed trend highlights the benefit of temporal averaging in mitigating the impact of atmospheric turbulence on piston estimation. As the number of averaged frames increases, the turbulence becomes increasingly smoothed out, allowing the differential piston signal to emerge more clearly from the speckle background. As discussed in \citet{Levraud:24} and \citet{Janin:24}, this behaviour is a direct consequence of the power spectral density ratio between the turbulence and the differential piston mode.
In the asymptotic limit, the RMSE is expected to decrease with the square root of the number of averaged frames (i.e.\ $\alpha_{lim} = -0.5$). However, we consistently measured slightly higher values. We attribute this deviation ($<6\%$ in all cases) to structural features in the PSD of the residual phase, which enhance the variance along the directions favored by the NN for differential piston estimation, thereby amplifying their realative influence. This ultimately leads to a slightly worse performance than expected.

\begin{table}[tb]
    \caption[]{List of parameters $\alpha$ and $\beta$ for the uncorrelated, correlated, and polychromatic configurations.}
    \label{tab:RMSE}
    \centering
    \begin{tabular*}{\linewidth}{@{\extracolsep{\fill}} ccc}
        &UNCORRELATED\\
        \hline
        \noalign{\smallskip}
        Strehl ratio & $\alpha$                    & $\beta$ (in nm)\\
        \hline
        \noalign{\smallskip}
        10\% & -0.471  & 106\\
        25\% & -0.472 & 74.3\\
        50\% & -0.480  & 49.6\\
        \hline
        \\

        &CORRELATED\\
        \hline
        \noalign{\smallskip}
        Strehl ratio & $\alpha$                    & $\beta$ (in nm)\\
        \hline
        \noalign{\smallskip}
        25\% & -0.389 & 105.8   \\
        \hline
        \\

        &POLYCHROMATIC\\
        \hline
        \noalign{\smallskip}
        Strehl ratio & $\alpha$                    & $\beta$ (in nm)\\
        \hline
        \noalign{\smallskip}
        25\% & -0.477 & 78.3\\
        \hline
    \end{tabular*}
\end{table}

We also investigated the relationship between Strehl ratio and RMSE. Figure~\ref{fig:uncorrelated_strehl} shows the variation of the $\beta$ parameter as a function of the Strehl ratio. As expected, higher Strehl ratios correspond to lower RMSE values, and a logarithmic dependency is observed between the two variables.

Finally, we derived a functional relationship linking RMSE to both the number of averaged frames and the Strehl ratio. Based on Eq.~\ref{eq:RMSE_n} and the logarithmic behaviour with respect to the Strehl ratio observed in Fig.~\ref{fig:uncorrelated_strehl}, we propose the following expression,
\begin{equation}
    \label{eq:RMSE_all}
    \mathrm{RMSE}(n_{uncorr},\,\mathrm{SR}) = \left( \gamma \cdot \ln(\mathrm{SR}) + \delta \right) \cdot n_{uncorr}^{\alpha}.
\end{equation}
Fitting this expression to all data points in Fig.~\ref{fig:uncorrelated} yields $\alpha \simeq -0.4680$, $\gamma \simeq -35.10$, and $\delta \simeq 186.5$.
This equation provides, in the case of uncorrelated turbulence, a predictive model for the fundamental RMSE limit achievable by our method.

\subsubsection{Correlated phasescreens}
\label{sec:results:temporal_averaging:correlated}

\begin{figure}[t]
    \centering
    \includegraphics[width=0.9\linewidth]{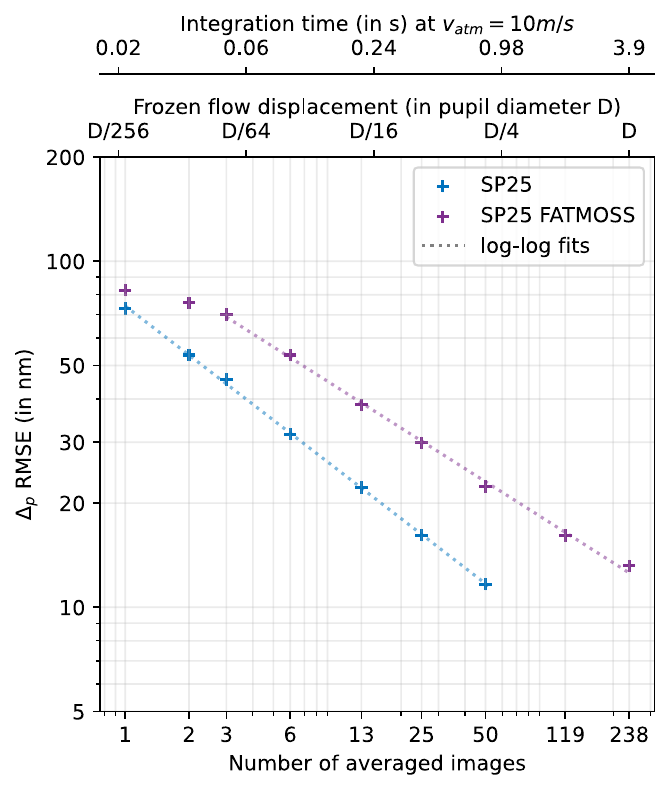}
    \caption{Influence of phasescreen simulation algortihm on differential piston estimation accuracy.
        Differential piston RMSE versus number of averaged images for the correlated SP25 FATMOSS condition (in purple). Results from the SP25 condition are plotted for reference (in blue, corresponding to SP25 in Fig.~\ref{fig:uncorrelated}).}
    \label{fig:uncorrelated_vs_fatmoss}
\end{figure}

\begin{figure}[t]
    \centering
    \includegraphics[width=0.9\linewidth]{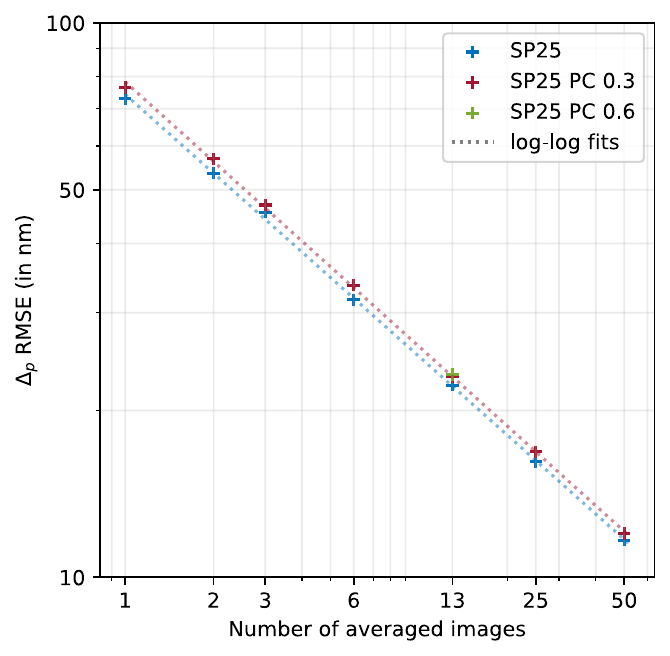}
    \caption{Influence of chromatism on differential piston RMSE.
        Differential piston RMSE versus number of averaged images for the polychromatic SP25 PC 0.3 (in dark red) and SP25 PC 0.6 (one data point, in green) configurations. Results from the SP25 condition are plotted for reference (in blue, corresponding to SP25 in Fig.~\ref{fig:uncorrelated}).
    }
    \label{fig:polychromatic}
\end{figure}

Building upon the results obtained with independent frames in the previous section, we now investigate the impact of a more realistic atmospheric model incorporating frozen flow with added boiling on the NN's performance.

We restricted this analysis to the SP25 FATMOSS dataset, as generating correlated phase screens requires significantly more averaged images to achieve comparable results and is computationally more demanding. Figure~\ref{fig:uncorrelated_vs_fatmoss} presents the differential piston RMSE as a function of the number of averaged frames. The results from the uncorrelated SP25 case (reported in blue, from Fig.~\ref{fig:uncorrelated}) serve as a reference baseline. As expected, for a given number of averaged frames, the RMSE values in the correlated case consistently exceed those in the uncorrelated configuration.

The upper axes of Fig.~\ref{fig:uncorrelated_vs_fatmoss} translate the number of averaged frames into equivalent frozen-flow displacements (in units of pupil diameter), as well as into on-sky integration times on the ELT, assuming a wind speed of $v_{\mathrm{wind}} = 10\,\mathrm{m/s}$. In our simulations, the frozen-flow was configured to shift by one pixel per frame to minimise the number of frames required for simulation, while ensuring sufficient temporal sampling. For instance, averaging 25 frames in the SP25 FATMOSS dataset results in an RMSE of approximately 30 nm. To achieve this, the frozen flow must move by one-quarter of the pupil diameter, corresponding to an integration time of 0.98 s at a wind speed of 10 m/s.
From the fit shown in Fig.~\ref{fig:uncorrelated_vs_fatmoss}, we can derive the following empirical relation:
\begin{equation}
    \label{eq:RMSE_corr}
    \begin{aligned}
        \mathrm{RMSE}(v_{\mathrm{atm}},\,t,\,\mathrm{SR}=25\%) & = \beta \cdot n_{corr}^\alpha \\&= \beta \cdot \left(\frac{d}{D} v_{\mathrm{atm}} \, t \right)^{\alpha},
    \end{aligned}
\end{equation}
where $n_{corr}$ is the number of correlated frames, which can be expressed in terms of $v_{atm}$ (the frozen-flow speed), $t$ (the integration time), $d$ (the simulated ELT pupil diameter), and $D$ (the ELT pupil diameter in real life). In this configuration, the fixed parameter values are $\beta\simeq 103.9~\mathrm{nm}$, $d=238~\mathrm{pixels,}$ and $D=39~\mathrm{m}$.

A small discrepancy is visible between the uncorrelated and correlated conditions for single-frame averaging. In principle, both configurations should yield identical results for $n = 1$, however, this was not observed for the present study. We attribute this difference to the presence of low spatial frequencies (i.e. below the 1 cycle per pupil cutoff captured by the DFT method of \citet{McGlamery:1976}) in the FATMOSS simulations, which are absent from the uncorrelated datasets. These low-frequency components degrade performance by reducing the signal-to-speckle ratio in the regions most critical for piston estimation, as discussed in \citet{Levraud:24} and \citet{Janin:24}.
After an initial transient regime, the FATMOSS results follow a consistent linear trend in the log-log scale, mirroring the behaviour observed for the uncorrelated case in Fig.~\ref{fig:uncorrelated}. The $\beta$ parameter is lower than in the uncorrelated case, but it does not have an absolute physical meaning. This slope is simply a direct consequence of the frozen-flow wind speed, the uncorrelated regime corresponding to the limiting case where the wind speed tends to infinity.

\subsection{Chromatism}
\label{sec:results:chromatism}

To evaluate the impact of chromatism on performance, we created the SP25 PC 0.3 dataset, which has a spectral bandwidth of 0.3~$\mu$m and shares the same central wavelength (1.65~$\mu$m) as SP25. We tested our algorithm using different numbers of averaged (uncorrelated) frames and the results are shown in Fig.~\ref{fig:polychromatic}.

As illustrated in the figure, the degradation in RMSE due to polychromatism is modest. The $\beta$ parameter (i.e. RMSE for single-frame averaging) increases slightly from approximately 74.3 nm in the monochromatic case to 78.3 nm in the polychromatic case (see Table~\ref{tab:RMSE}). Since the log-log slopes are equivalent for both regimes, the RMSE ratio between the monochromatic and polychromatic cases remains nearly constant, representing a difference of approximately 5\% over the entire range of averaging values.

\begin{figure}
    \centering
    \includegraphics[width=.9\linewidth]{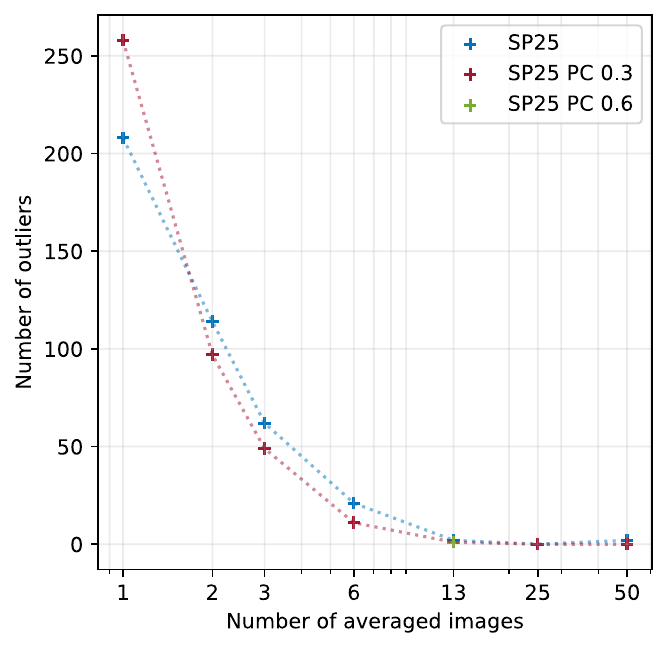}
    \caption{Influence of chromatism on outliers mitigation.
        Number of outliers versus number of averaged images for the SP25 (blue), SP25 PC 0.3 (dark red), and SP25 PC 0.6 (one data point, green) conditions.
    }
    \label{fig:polychromatic_outliers}
\end{figure}

As discussed in Section~\ref{sec:method:pi_ambiguity}, polychromatism can also be exploited to increase the capture range of piston sensors by introducing a finite coherence length.
We further investigated whether polychromatism mitigates outliers, based on the hypothesis that a finite coherence length could help resolve $2\pi$ ambiguities leading to large piston errors in monochromatic conditions. To test this, we computed the number of outliers as a function of the number of averaged frames. The results, presented in Fig.~\ref{fig:polychromatic_outliers}, show that polychromatic data consistently yield fewer outliers, except in the single-frame case. However, the overall improvement remains limited and does not provide a clear advantage in the polychromatic case.

In conclusion, polychromatism does not significantly degrade the performance of differential piston estimation, causing less than a 5\% increase in RMSE; therefore, it is not a limiting factor. It provides only marginal benefits in reducing the occurrence of outliers in its current implementation. However, there is potential for further improvement in this area.

\subsection{Detector noise}
\label{sec:results:noise}

\begin{figure}
    \centering
    \includegraphics[width=.9\linewidth]{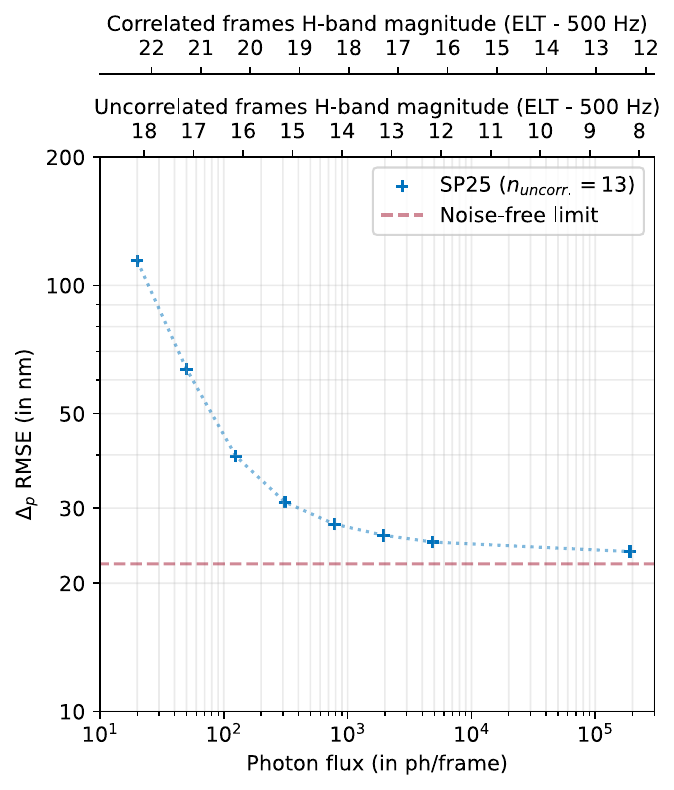}
    \caption{Influence of photon count on differential piston estimation accuracy.
        Differential piston RMSE versus number of photons for the SP25 turbulence configuration, with 13-frame averaging. Red dotted line represents the noise-free limit from 13-frame averaged SP25 in Fig.~\ref{fig:uncorrelated}.
        Top axes show the corresponding source magnitude, assuming an ELT-like system with a wavefront sensor operating at 500Hz. The lower top axis corresponds to 13-frame averaging under the uncorrelated regime, while the upper top axis corresponds to the effective integration time of $\sim1$s in real-life operation (see Sect.~\ref{sec:results:noise} for details).
    }
    \label{fig:noise}
\end{figure}

To assess the robustness of our method against detector noise, we introduced both photon noise and a readout noise of $1\,\mathrm{e}^{-}$. We evaluated the performance under the SP25 condition at $n_{uncorr}=13$ frames, on average. We varied the number of input photons and measure the resulting RMSE of the differential piston estimates. Figure~\ref{fig:noise} shows the RMSE as a function of the number of photons per frame in the uncorrelated SP25 configuration averaged over 13 frames. As expected, the RMSE increases as the photon count decreases. For reference, the noise-free RMSE limit (previously shown in Fig.~\ref{fig:uncorrelated}) is overlaid as a red dotted line.

The top axes of Fig.~\ref{fig:noise} show the corresponding source magnitude, assuming an ELT-like system with a wavefront sensor operating at 500Hz. The lower top axis corresponds to 13-frame averaging using uncorrelated sub-PSFs. However, these magnitudes do not directly apply to real-life observing conditions, where atmospheric turbulence is temporally correlated, and more frames are needed to achieve the same RMSE level.
Following Eqs.~\ref{eq:RMSE_n} and \ref{eq:RMSE_corr}, we have
\begin{equation}
    n_{\mathrm{corr}} = \left( \frac{\beta_{\mathrm{uncorr}}}{\beta_{\mathrm{corr}}} \, n_{\mathrm{uncorr}}^{\alpha_{\mathrm{uncorr}}} \right)^{1/\alpha_{\mathrm{corr}}}\,.
\end{equation}
Using the values from Table~\ref{tab:RMSE}, we found 13 uncorrelated frames correspondng to approximately 56 correlated frames, in agreement with Fig.~\ref{fig:uncorrelated_vs_fatmoss}.
The conversion between the number of simulated frames and the corresponding real integration time, given the simulation parameters, is expressed as
\begin{equation}
    t = \frac{D}{238} \, \frac{n_{\mathrm{corr}}}{v_{\mathrm{atm}}}\,.
\end{equation}
This implies that 13 uncorrelated frames correspond to an integration time of $t \simeq 0.91~\mathrm{s}$ for $v_{\mathrm{atm}} = 10~\mathrm{m/s}$.
At a frame rate of 500 Hz, this translates to approximately 450 individual camera frames.
Therefore, to reflect realistic observing conditions, the magnitudes shown in Fig.~\ref{fig:noise} for 13 uncorrelated frames must be corrected for this increased integration time. This correction translates into an effective gain of approximately $(450/13)^{1/2.5} \simeq 4.1$ magnitudes, as indicated by the upper top axis.

\begin{figure}
    \centering
    \includegraphics[height=.9\linewidth]{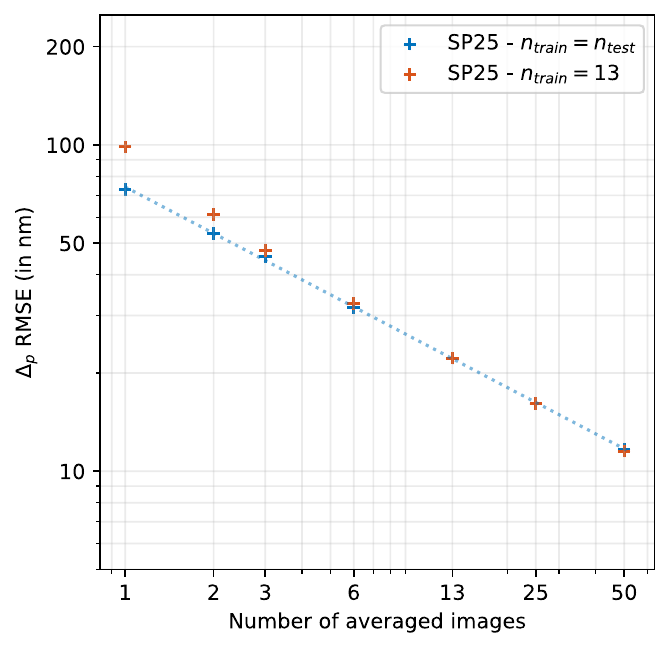}
    \caption{Performance of a NN trained at fixed image count under varying averaging number.
        Differential piston RMSE versus number of averaged images for the SP25 turbulence configuration, as estimated by a NN trained on 13-frame averaged images.
        The results from the SP25 condition are plotted for reference (in light blue, corresponding to SP25 in Fig.~\ref{fig:uncorrelated}).}
    \label{fig:crosstest_average}
\end{figure}

For a realistic operational case (i.e. assuming a correlated regime), the 18\textsuperscript{th} limiting magnitude of the HARMONI LTAO mode yields a RMSE of approximately 27.5 nm for an arbitrary 1 second integration, which is only 5.3 nm higher than the noiseless situation. This suggests that the primary limitation to performance is not photon flux, but rather the limited effectiveness of speckle averaging, as discussed in Sect.~\ref{sec:results:temporal_averaging}. To elaborate on this point, we note from Fig.~\ref{fig:noise} that the RMSE in the presence of noise follows the following empirical relation,
\begin{equation}
    \label{eq:RMSE_noise}
    \mathrm{RMSE} \propto \exp \left(1/n_{\mathrm{uncorr}}\right).
\end{equation}
The decay rate of the RMSE described by Eq.~\ref{eq:RMSE_noise} is slower than that of the noiseless case (Eq.~\ref{eq:RMSE_n}), which can be formally expressed as
\begin{equation}
    \label{eq:RMSE_decay}
    \frac{\partial}{\partial n}\mathrm{RMSE}\,\big\rvert\,_{\text{speckle}} < \frac{\partial}{\partial n}\mathrm{RMSE}\,\big\rvert\,_{\text{noise}}\,,
\end{equation}
reinforcing the conclusion that residual speckle patterns, rather than photon noise, constitute the main limiting factor in the differential piston estimation.

We note that the 1~s timescale was chosen arbitrarily. If the piston dynamics were significantly faster than this value, the method would still remain applicable, albeit with a degraded final RMSE. This is in line with the trends discussed in Section~\ref{sec:results:temporal_averaging}.

\subsection{Robustness analysis}
\label{sec:results:crosstests}

In this section, we quantify the robustness of our NN to mismatches between training and inference conditions. Specifically, we assess its sensitivity to (1) the number of averaged frames (Section~\ref{sec:results:crosstests:average}); (2) the Strehl ratio (Section~\ref{sec:results:crosstests:strehl}); and (3) detector noise (Section~\ref{sec:results:crosstest:noise}).

\begin{figure}[t]
    \centering
    \includegraphics[height=.9\linewidth]{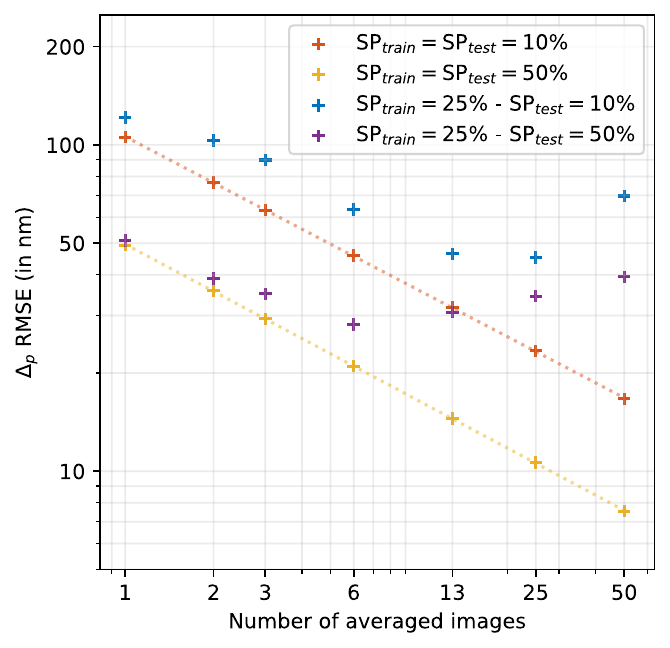}
    \caption{Influence of training the Strehl ratio on a differential piston estimation accuracy compared to training on Strehl single values.
        Differential piston RMSE versus number of averaged images for the SP10 and SP50 turbulence configurations, as estimated by a NN trained on the SP25 condition.
        Results from the tailored network are plot for reference (light red and light yellow for SP10 and SP50 respectively from Fig.~\ref{fig:uncorrelated}).}
    \label{fig:crosstest_SP_single}
\end{figure}

\begin{figure}[t]
    \centering
    \includegraphics[width=.9\linewidth, trim=0 0 0 -5, clip]{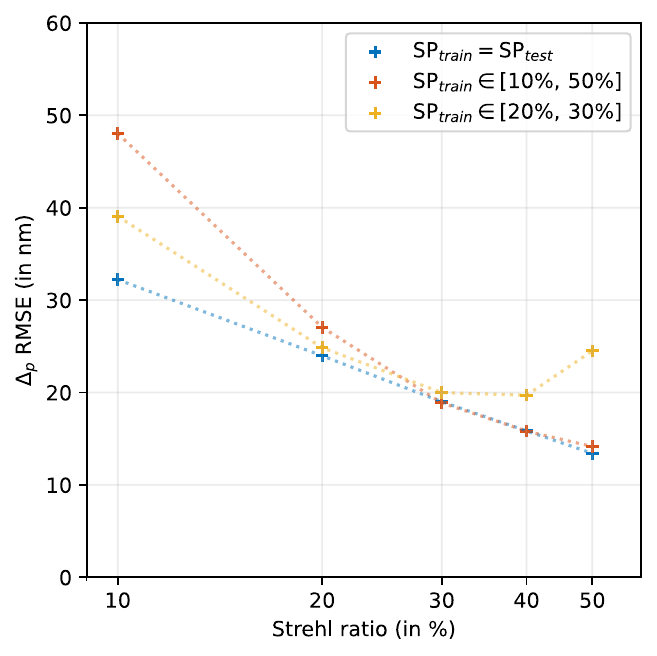}
    \caption{Influence of training the Strehl ratio on a differential piston estimation accuracy compared to training on Strehl ranges.
        Differential piston RMSE versus Strehl ratio for networks trained on the 10-50\% range (red) and 20-30\% (yellow). Results from the tailored network are plotted in blue for reference.}
    \label{fig:crosstest_SP_range}
\end{figure}

\subsubsection{Averaging}
\label{sec:results:crosstests:average}

We present the results obtained using the SP25 13-frame averaging NN introduced in Section~\ref{sec:results:temporal_averaging} and evaluate its performance across varying numbers of averaged frames. Figure~\ref{fig:crosstest_average} displays the results, with the reference curve from Fig.~\ref{fig:uncorrelated} included in light blue for comparison.

For numbers of averaged frames greater than 13, the NN performs at the same level as a dedicated NN trained specifically for that averaging level. However, for fewer than 13 frames, the discrepancy increases as the number of frames decreases, reaching a difference of nearly 30 nm in RMSE for the single-frame case. This indicates that even when the NN is trained on a single averaging value, it achieves relatively good performance close to the theoretical limit over a wide range of averaging values. Therefore, this parameter does not require fine tuning, as the NN demonstrates robustness to its variation.

\subsubsection{Strehl ratio}
\label{sec:results:crosstests:strehl}

We went on to perform a similar test, this time focusing on the impact of the Strehl ratio. Specifically, we used the networks trained on the SP25 dataset for each averaging values and evaluate their performance on the SP10 and SP50 datasets. The results are presented in Fig.~\ref{fig:crosstest_SP_single}, where the reference performances for SP10 and SP50 are shown in light red and yellow, respectively.

In this case, the RMSE difference increases with the number of averaged frames. This indicates that networks trained at 25\% Strehl lose more performance as averaging increases when applied to datasets with mismatched Strehl ratios. In other words, variations in Strehl ratio appear to be more detrimental than changes in the number of averaged frames. This result is particularly concerning for real-world applications, where the Strehl ratio can vary significantly during observations.

To further investigate this effect, we trained two additional networks on different Strehl ranges and compared their performance. Figure~\ref{fig:crosstest_SP_range} shows the results for these two networks: one trained on the narrower Strehl range $[20\%, 30\%]$, and the other on the broader range $[10\%, 50\%]$.

The $[20\%, 30\%]$ NN performs well on the 20\% and 30\% cases, showing only a few nanometres of RMSE difference compared to dedicated networks. Outside this range, the error increases marginally, reaching approximately 10 nm for a Strehl ratio of 50\%, which represents a clear improvement over the results obtained earlier with the NN trained solely at 25\% Strehl (see Fig.~\ref{fig:crosstest_SP_single}).
The $[10\%, 50\%]$ NN performs better than the $[20\%, 30\%]$ NN in the 25\%–50\% range, with only a fraction of a nanometre RMSE difference compared to the tailored networks. For 10\% and 20\% Strehl, however, the RMSE difference increases to several tens of nanometres.
Overall, the $[20\%, 30\%]$ NN outperforms the $[10\%, 50\%]$ NN for Strehl values between 10\% and approximately 25\%, while the $[10\%, 50\%]$ NN performs better above 25\%.

The fact that the $[20\%, 30\%]$ NN performs better than the $[10\%, 50\%]$ one in the 10\%–20\% Strehl range may be explained by the distribution of Strehl ratios in the training data. The $[10\%, 50\%]$ NN is exposed to a broader range of values, and its training may converge toward a solution that favours higher Strehl ratios in order to minimise the overall loss, resulting in degraded performance at the lower end of the range.

\begin{figure}
    \centering
    \includegraphics[width=.9\linewidth]{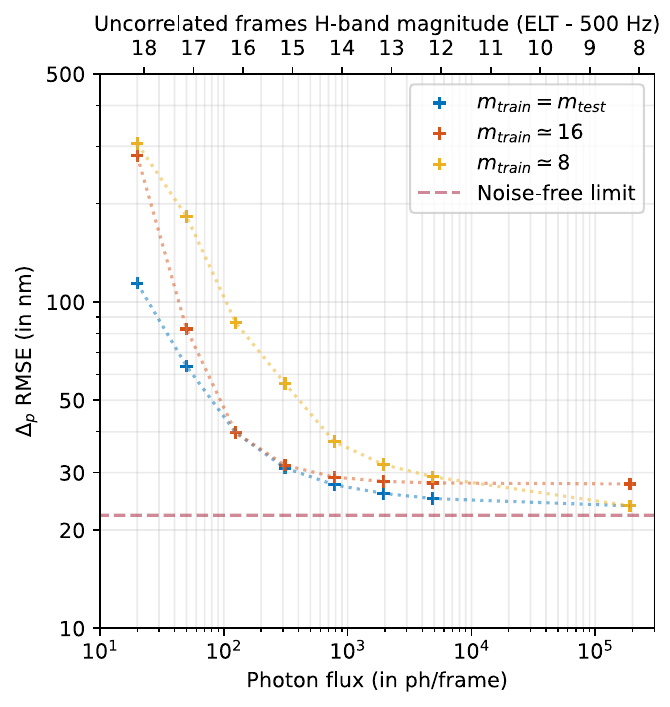}
    \caption{Influence of training the magnitude on a differential piston estimation accuracy compared to training on a single magnitude value.
        Differential piston RMSE versus number of photons for the SP25 turbulence configuration, with 13-frame averaging.
        Results are shown for three regimes within a monochromatic uncorrelated setting: the baseline model (blue, corresponding to Fig.~\ref{fig:noise}), a model trained at magnitude $m = 16.5$ (red), and a model trained at $m = 13.5$ (yellow).
        The dashed purple line indicates the noise-free monochromatic limit.
    }
    \label{fig:crosstest_noise_single}
\end{figure}

\begin{figure}[ht]
    \centering
    \includegraphics[width=.9\linewidth]{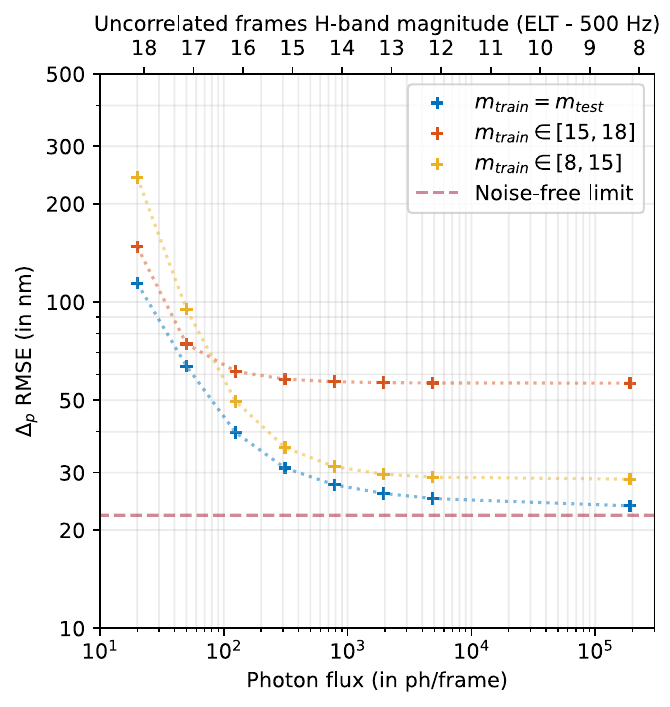}
    \caption{Influence of training the magnitude on a differential piston estimation accuracy compared to training on a magnitude range.
        Differential piston RMSE versus number of photons per frame for the SP25 turbulence configuration, with 13-frame averaging.
        Results are shown for three regimes within a monochromatic uncorrelated setting: the baseline model (blue, corresponding to Fig.~\ref{fig:noise}), a model trained at magnitude $m \in [15;18]$ (red), and a model trained at $m \in [8;15]$ (yellow).
        The dashed purple line indicates the noise-free monochromatic limit.}
    \label{fig:crosstest_noise_range}
\end{figure}

\subsubsection{Noise}
\label{sec:results:crosstest:noise}

To conclude this section on cross-testing, we present the results we obtained when considering noise. For this analysis, we first predicted the differential piston across a range of magnitudes using two networks trained on images at magnitudes 8 and 16, respectively. The results are shown in Figure~\ref{fig:crosstest_noise_single}. As expected, each NN performs best near its respective training magnitude, approaching the theoretical limit derived from Figure~\ref{fig:noise}, which is shown in light blue for reference.

The NN trained on magnitude 16 outperforms the one trained on a magnitude of 8 for magnitudes greater than approximately 12. At lower magnitudes, both networks converge toward different limits: the NN trained on a magnitude of 8 is aligned with the theoretical lower bound, while the NN trained on magnitude 16 performs worse by few nanometers.

In line with the Strehl ratio analysis in Section~\ref{sec:results:crosstests:strehl}, we also evaluated two additional networks trained over magnitude ranges [8, 15] and [15, 18]. The results are shown in Fig.~\ref{fig:crosstest_noise_range}. The [8, 15] NN maintains strong performance across a wide magnitude range, staying close to the theoretical limit. The consistent separation between the two curves in the logarithmic scale suggests a constant RMSE ratio of approximately 8\%.
For the [15, 18] NN, performance is satisfactory for magnitudes above 17, but rapidly saturates below this value, stabilising at an RMSE of approximately 60 nm.

As stated in Sect.~\ref{sec:results:noise}, theses results were computed from the uncorrelated SP25 datasets and need to be converted to correlated magnitudes to draw conclusions applicable to a real atmosphere. Secondary magnitude axis is recalled at the top for ease of translation.
These behaviours are reinsuring in the sense that the NN trainend on the [8, 15] magnitude range is capable of outputing reasonable performance all along the considered magnitudes.

\section{Applications to other type of PSF}
\label{sec:applications}

\begin{figure}
    \centering
    \begin{subfigure}[b]{.49\linewidth}
        \centering
        {\includegraphics[width=.9\linewidth, trim=100 40 100 40, clip]{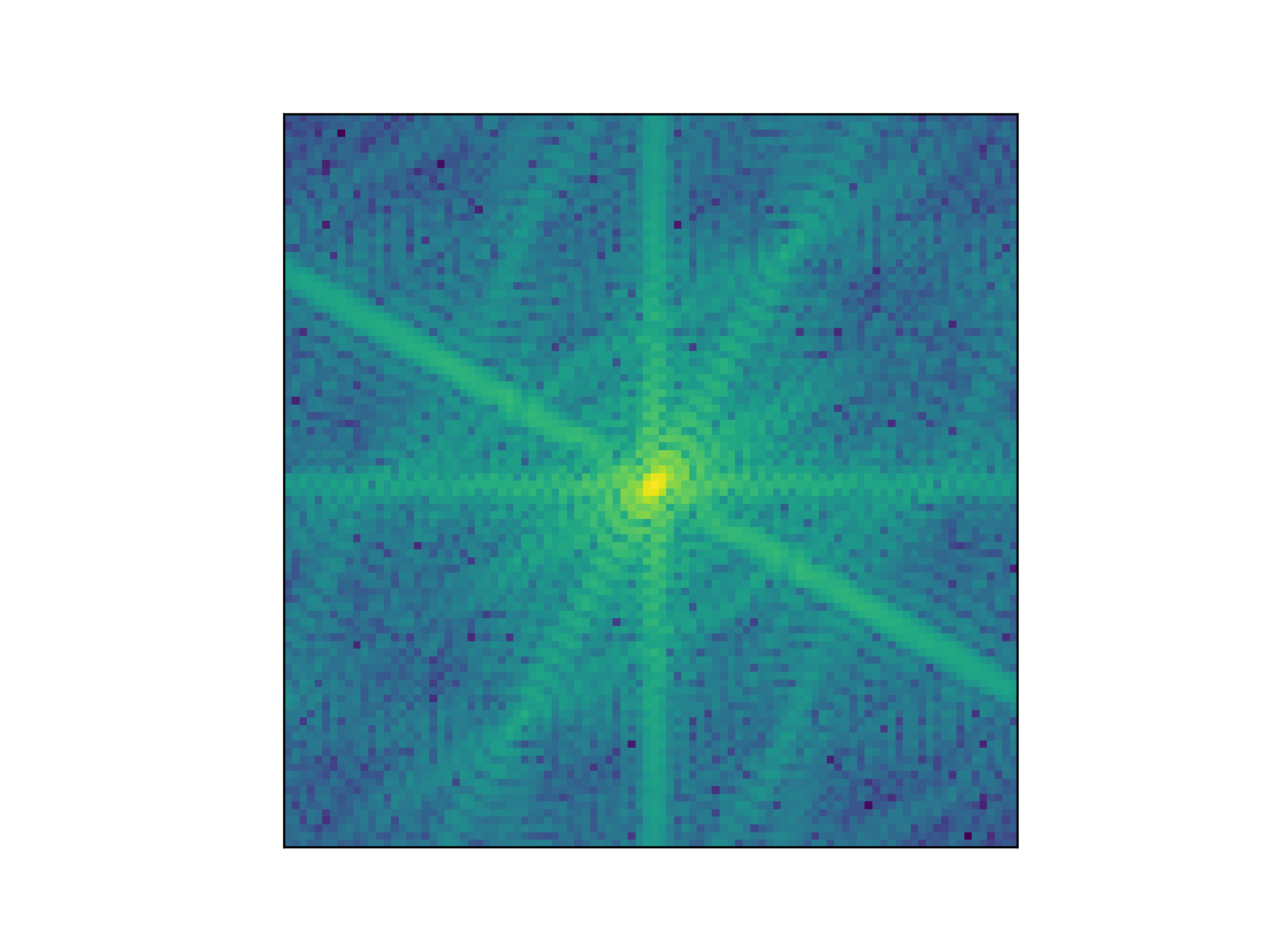}}
        \caption{}
        \label{fig:LIFT:a}
    \end{subfigure}%
    \begin{subfigure}[b]{.49\linewidth}
        \centering
        {\includegraphics[width=.9\linewidth, trim=100 40 100 40, clip]{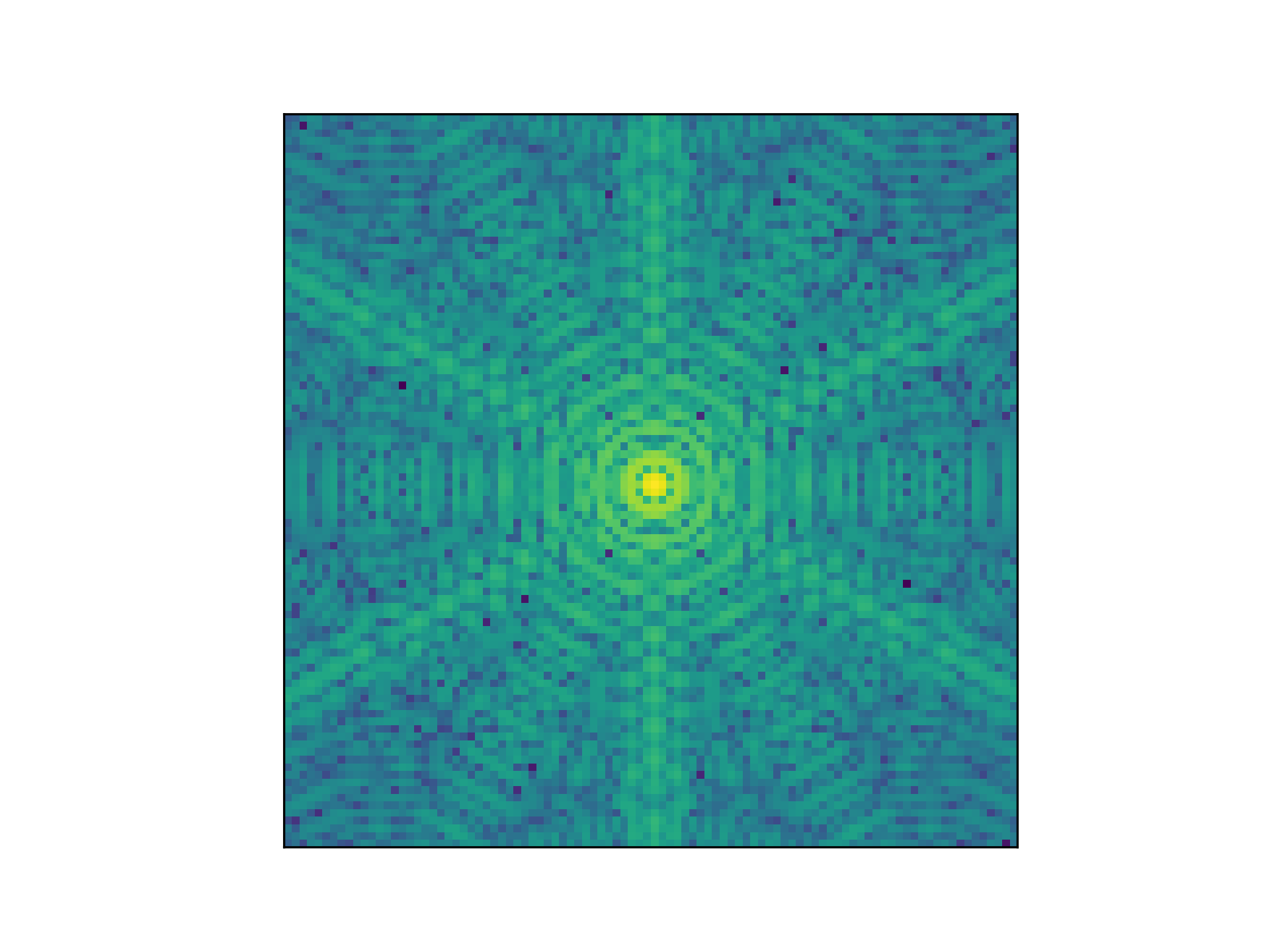}}
        \caption{}
        \label{fig:LIFT:b}
    \end{subfigure}
    \begin{subfigure}[b]{.49\linewidth}
        \centering
        {\includegraphics[width=.9\linewidth, trim=100 40 100 40, clip]{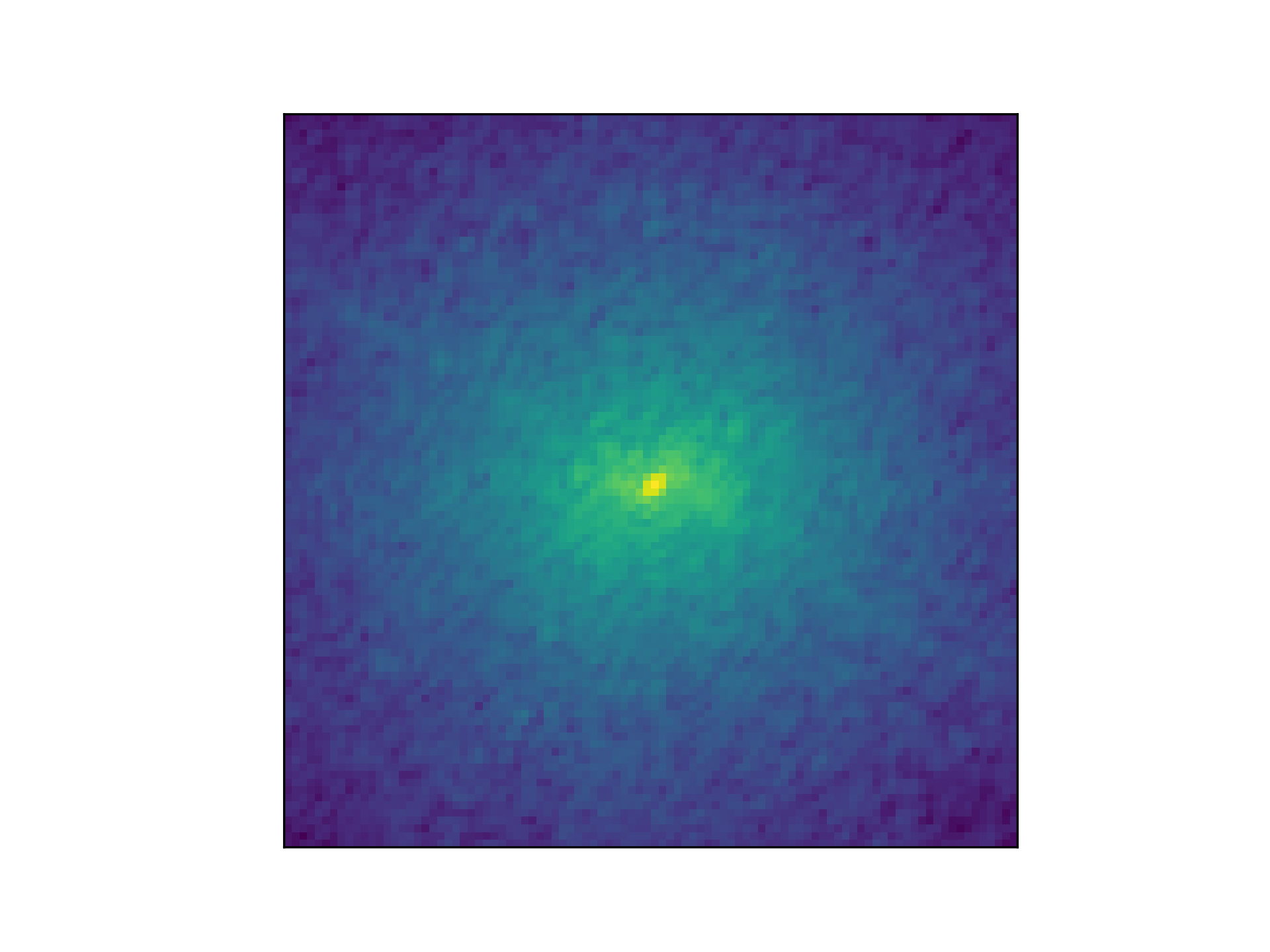}}
        \caption{}
        \label{fig:LIFT:c}
    \end{subfigure}
    \begin{subfigure}[b]{.49\linewidth}
        \centering
        {\includegraphics[width=.9\linewidth, trim=100 40 100 40, clip]{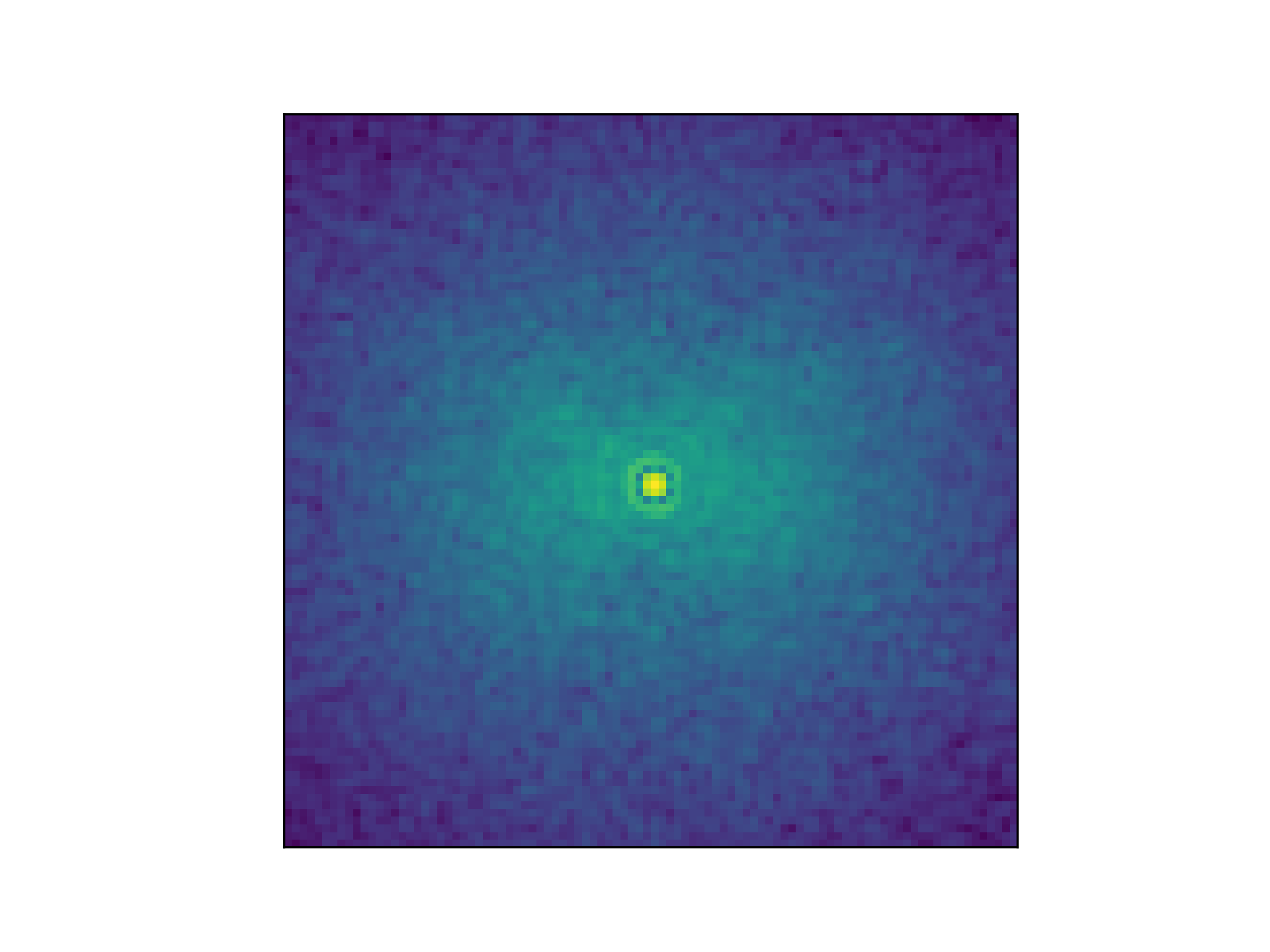}}
        \caption{}
        \label{fig:LIFT:d}
    \end{subfigure}
    \caption{
        \label{fig:LIFT}
        Top: Diffraction-limited PSFs for the (a) UR subpupil of the SH-WFS and (b) LIFT configurations.
        Bottom: Long-exposure PSFs under the SP50 condition for the (c) UR subpupil of the SH-WFS and (d) LIFT configurations.}
\end{figure}

For the sake of completeness, we briefly discuss the applicability of our method to systems producing PSFs other than the one produced by the $2\times2$ SH-WFS configuration presented throughout this paper.
The NN architecture and training procedure described in Sect.~\ref{sec:method:nn_architecure} are designed to generalise to various PSF types, provided that the input variables remain consistent with those discussed throughout this paper.
To illustrate this, we tested the NN on full pupil LIFT-like PSFs. For reference, LIFT~\citep{Plantet:13} is a focal-plane wavefront sensor that iteratively estimates phase aberrations in presence of a known phase offset (typically a defocus or astigmatism).
This optical configuration produces PSFs with distinct spatial characteristics while preserving the same differential piston encoding behavior, making it a valuable test case for assessing the robustness and versatility of our approach. Figure~\ref{fig:LIFT} presents the PSFs for the upper-right (UR) subpupil in both the SH-WFS and LIFT configurations, shown in the diffraction-limited case (left) and under the SP50 condition (right).

We generated a new dataset of full-pupil PSFs (sampled at the Shanon limit) with an additional defocus phase of 1 radian under the SP25 13-frame averaging conditions.
The dataset was produced following the same procedure described in Sect.~\ref{sec:method:training}. For this test, neither noise nor polychromatic effects were included, consistent with the set-up used for the results presented in Sect.~\ref{sec:results:temporal_averaging:correlated}.
Random piston is introduced on all segments of M4. While the NN is capable of inferring multiple pistons from the same image~\citep[Sect. 5.4]{Janin:24}, we trained it only to retrieve the differential piston between segments 2 and 3, as done in all previous sections of this paper, for simplicity. The NN is expected to perform at the same level in a configuration where it must retrieve all six differential pistons.

We obtained a RMSE of approximately 20.3 nm, which compares favorably to the 22.2 nm achieved on Fig.~\ref{fig:uncorrelated} with the $2\times2$ sub-PSFs. This result is encouraging and suggests that the NN can be effectively adapted to alternative focal plane WFS, such as LIFT, with minimal modifications to the training procedure.

\section{Conclusions}
\label{sec:conclusions}

In this work, we demonstrate that a NN can reliably estimate differential piston from $2\times2$ SH-WFS sub-PSF images. We evaluated the performance limits of our method under a range of conditions, including temporal averaging (Sect.~\ref{sec:results:temporal_averaging}), chromatic effects (Sect.~\ref{sec:results:chromatism}), noise (Sect.~\ref{sec:results:noise}), and robustness to discrepency between the training and testing phase (Sect.~\ref{sec:results:crosstests}).

From noiseless simulations, we derived an empirical relation that links the RMSE to both the number of averaged frames and the Strehl ratio. This model can be used to predict the expected RMSE given specific observation conditions.
We then extended our analysis to more realistic turbulence scenarios by considering correlated atmospheric effects. We found that the power-law behavior with respect to the number of averaged frames still holds with different parameter values after a short transition regime. This indicates that frame averaging in real-life conditions follows the same trend as in the uncorrelated case, but requires a larger number of frames to reach a comparable RMSE that mainly depends on the frozen-flow windspeed.

Chromatic effects were found to have minimal impact, with RMSE increasing by less than 5\% compared to the monochromatic case. This confirms that polychromatic light can be used without performance degradation and potentially even leveraged to increase the sensor’s capture range in future developments. While noise does affect the RMSE, we have shown that its impact is secondary compared to that of speckle averaging (Eq.~\ref{eq:RMSE_decay}), indicating that photon noise combined with a read-out noise of $1\,\mathrm{e}^{-}$ is not the primary factor limiting performance.

Considering the capture range, we have shown that our method can consistently measure differential piston over a range approaching $\pi$ radians, which corresponds to nearly 800 nm in the H band. This range should be sufficient to address the levels of piston induced by the LWE expected on the ELT. For short bursts of AO-induced petal modes, which typically last a few milliseconds, the network may still be able to measure them, albeit with a higher RMSE. A hybrid strategy could be implemented to address both cases, either by relying on long integration times for LWE correction alone or by using short integrations to capture fast transient events, while still enabling LWE correction through the summation of the short-exposure frames. The expected performance of such strategies remains to be quantified.

Finally, we validated that our network architecture is robust across a range of conditions. Training under one configuration and evaluating under another showed that the network was able to maintain a consistent performance. This relaxes the requirement for precisely matched training conditions and opens the door to more flexible and practical deployment strategies.

As an attempt at generalising our results, we extended the application of our network to a different type of input by testing it on LIFT images, with encouraging results. This provides a first indication of the method’s versatility and suggests that it may be adaptable to other focal-plane wavefront sensors with minimal modifications to the training procedure.

Moreover, while this study focuses on the $2\times2$ Shack-Hartmann wavefront sensor, selected for its photon efficiency and consistency with existing systems (e.g. HARMONI and MICADO), the proposed method is inherently generalisable to alternative SH-WFS geometries, including higher-order configurations such as $3\times3$ or $4\times4$. We anticipate a comparable performance in these cases, as the underlying NN framework is agnostic to the specific subaperture arrangement. The $2\times2$ configuration, however, serves as the baseline for this work due to its simplicity, robustness against detector noise and established role in low-order wavefront sensing.

\begin{acknowledgements}
    This work benefited from the support of the BPI with France2030 SSA-DOSSA project, the French National Research Agency (ANR) with APPLY (ANR-19-CE31-0011) and LabEx FOCUS (ANR-11-LABX-0013), the Programme Investissement Avenir F-CELT (ANR-21-ESRE-0008), the Action Spécifique Haute Résolution Angulaire (ASHRA) of CNRS/INSU co-funded by CNES, the ECOS-CONYCIT France-Chile cooperation (C20E02), the ORP-H2020 Framework Programme of the European Commission’s (Grant number 101004719), STIC AmSud (21-STIC-09), the french government under the France 2030 investment plan, the Initiative d’Excellence d’Aix-Marseille Université A*MIDEX, program number AMX-22-RE-AB-151. This research has made use of computing facilities operated by CeSAM data center at LAM, Marseille, France.
    Pierre Janin-Potiron warmly thanks Mahé Janin-Portella for insightful discussions during the revision process.
\end{acknowledgements}

\bibliographystyle{bibtex/aa}
\bibliography{aa56704-25.bib}

%
%




\end{document}